\documentclass[aps, prd,floats, floatfix, twocolumn,superscriptaddress, nofootinbib, showpacs]{revtex4}

\usepackage{graphicx}
\usepackage{hyperref}
\usepackage{amsmath,amssymb}
\usepackage{amsfonts}
\usepackage{xspace} 
\usepackage[usenames]{color}
\usepackage{dcolumn} 
\usepackage{bm} 
\usepackage{epsfig}
\usepackage{upgreek}

\usepackage{ulem}
\definecolor {darkgreen}{rgb}{0.2,0.7,0.2}

\newcommand\be{\begin{equation}}
\newcommand\ba{\begin{eqnarray}}
\newcommand\ee{\end{equation}}
\newcommand\ea{\end{eqnarray}}

\newcommand{\mb}[1]{\mbox{\boldmath $#1$}}

\newcommand{\I}{{\cal{I}}}

\newcommand{\CS}{{\mbox{\tiny CS}}}
\newcommand{\QG}{{\mbox{\tiny QG}}}

\newcommand{\BK}{{\mbox{\tiny BK}}}
\newcommand{\SBK}{{\mbox{\tiny SBK}}}
\newcommand{\DK}{{\mbox{\tiny DK}}}

\newcommand{\MAT}{{\mbox{\tiny Mat}}}

\begin{document}
\title{Bumpy Black Holes in Alternative Theories of Gravity}
 
\author{Sarah Vigeland}
\affiliation{Department of Physics and MIT Kavli Institute, Cambridge, MA 02139, USA.}

\author{Nicol\'as Yunes}
\affiliation{Department of Physics and MIT Kavli Institute, Cambridge, MA 02139, USA.}

\author{Leo C.\ Stein}
\affiliation{Department of Physics and MIT Kavli Institute, Cambridge, MA 02139, USA.}

\date{\today}

\begin{abstract} 

We generalize the bumpy black hole framework to allow for alternative theory deformations. 
We construct two model-independent parametric deviations from the Kerr metric: one built from a generalization of the quasi-Kerr and bumpy metrics and one built directly from perturbations of the Kerr spacetime in Lewis-Papapetrou form. 
We find the conditions that these ``bumps'' must satisfy for there to exist an approximate second-order Killing tensor so that the perturbed spacetime still possesses three constants of the motion  (a deformed energy, angular momentum and Carter constant) and the geodesic equations can be written in first-order form. 
We map these parameterized metrics to each other via a diffeomorphism and to known analytical black hole solutions in alternative theories of gravity.
The parameterized metrics presented here serve as frameworks for the systematic calculation of extreme mass-ratio inspiral waveforms in parameterized non-GR theories and the investigation of the accuracy to which space-borne gravitational wave detectors can constrain such deviations.

\end{abstract}



\maketitle

\section{Introduction}
\label{intro}

Gravitational waves (GWs) will be powerful tools for learning about source astrophysics and testing strong field gravity~\cite{Schutz:2009tz,2010GWN.....4....3S}.  The Laser Interferometer Space Antenna (LISA)~\cite{Bender:1998,Danzmann:2003tv,Danzmann:2003ad,Prince:2003aa}, for example, is expected to be sensitive to roughly a few radians or better of GW phase in a one-year observation. Such levels of precision are achieved through matched filtering, where the data is cross-correlated with a family of waveform models (see eg.~\cite{lrr-2005-3}). If a signal is present in the data, this cross-correlation acts as a filter that selects the member of the waveform family that most closely resembles the signal. 

Extreme mass-ratio inspirals (EMRIs) are ideal astrophysical sources to perform precision GW tests of strong field gravity with LISA~\cite{AmaroSeoane:2007aw,Schutz:2009tz}. These sources consist of a small compact object (SCO) that spirals into a supermassive black hole (SMBH) in a generic orbit, producing millions of radians in GW phase inside LISA's sensitivity band. These waves carry detailed information about the spacetime geometry in which the SCO moves, thus serving as a probe to test general relativity (GR).  

Tests of strong field gravity cannot rely on waveform families that assume GR is correct {\emph{a priori}}. Instead, such tests must employ more generic waveforms that allow for GR deviations. Recently, Yunes and Pretorius~\cite{Yunes:2009ke} proposed the parameterized post-Einsteinian (ppE) framework, in which analytic waveforms that represent comparable mass-ratio coalescences in GR are parametrically deformed. For specific values of the ppE parameters, one recovers GR waveforms; otherwise they describe waveforms in non-GR theories. The ppE scheme has been shown to be sufficiently flexible to map all known alternative theory predictions for comparable mass-ratio coalescences~\cite{Yunes:2009ke}.

PpE deformations of EMRI waveforms have not yet been constructed because of the complexities associated with the computation of such waveforms. EMRIs are not amenable to post-Newtonian (PN) approximation schemes~\cite{Blanchet:2002av,Yunes:2008tw,2009GWN.....2....3Y,Yunes:2009ef,Yunes:2010zj},  from which many GR templates are today analytically constructed. Instead, to build EMRI waveforms one must solve geodesic-like equations, enhanced with a radiation-reaction force that induces an inspiral. The solution of such differential equations can only be found numerically (see eg.~\cite{Barack:2009ux} for a recent review of the self-force problem).  

One can parametrically deform EMRI waveforms by introducing non-GR deviations in the numerical scheme used to build such waveforms. Two main ingredients can then be modified: the conservative sector, which controls the shape of non-radiative orbits, and the dissipative sector, which controls the rate of inspiral and the GW generation mechanism. The conservative sector, to leading order in the mass-ratio, depends only on the {\emph{background}} spacetime metric, the Kerr metric, on which the SCO moves like a test-particle.

The goal of this work is to find a parametric deformation of the Kerr metric that allows for non-GR deviations while retaining a {\emph{smooth Kerr limit}}, i.e.~as the deformation parameters go to zero, the deformed Kerr metric reduces smoothly and exactly to Kerr. This limit then guarantees that the main properties of the Kerr background survive to the deformation, such as the existence of an event horizon, an ergosphere, and, in particular, three constants of the motion, such that the geodesic equations can be separated into first-order form.  

Although motion in alternative theories of gravity need not be geodesic, we will here assume that the geodesic equations hold to leading order in the mass ratio. This then allows us to focus only on metric deformations that correct the conservative sector of EMRI waveforms. This assumption is justified for theories that derive from certain diffeomorphism-covariant Lagrangians~\cite{2010PhRvD..81h4060G}, as is the case, for example, in dynamical Chern-Simons (CS) modified gravity~\cite{Alexander:2009tp}, in Einstein-Dilaton-Gauss-Bonnet theory~\cite{Kanti:1995vq,Torii:1996yi,Kanti:1997br,Pomazanov:2003wq,Pani:2009wy}, and in dynamical quadratic gravity~\cite{1992PhLB..285..199C,Mignemi:1992pm,Yunes:2011we}. We are not aware of any alternative theory where this is not the case to leading order in the mass ratio (and neglecting spins), although it is not hard to imagine that one may exist.

A framework already exists to parametrically deform the metric tensor through the construction of so-called {\emph{bumpy spacetimes}}~\cite{Collins:2004ex,2010PhRvD..81b4030V,2010arXiv1008.1278V}. Such metrics are deformations of the Kerr metric, where the bumps are required to satisfy the Einstein equations. Because of this last condition, bumpy black holes (BHs) are not sufficiently generic to represent BHs in certain alternative theories of gravity (e.g.~solutions that are not Ricci flat). A better interpretation is to think of these bumps as representing exterior matter distributions. The bumpy BH formalism then allows tests of whether compact objects are truly described by the {\emph{vacuum}} Kerr metric or by some more general tensor with external matter sources, assuming GR still holds. 

We here propose two generalizations of the bumpy BH formalism to allow for metric deformations that can represent vacuum BH solutions in alternative theories. The first approach, the generalized bumpy Kerr (BK) scheme, takes the standard bumpy metric and relaxes the requirement that the bumps satisfy the Einstein equations. The second approach, the generalized deformed Kerr (DK) metric, perturbs the most general stationary, axisymmetric metric in Lewis-Papapetrou form and transforms it to Boyer-Lindquist coordinates without assuming the deformations satisfy the Einstein equations. In both cases, the metric is exactly Kerr to zeroth order in the deformation parameter, but it deviates from Kerr at first order through a set of arbitrary functions. 

The metric deformation is then restricted by requiring that the full metric still possesses three constants of the motion: a conserved energy (associated with time invariance), a conserved angular momentum component (associated with azimuthal rotational invariance), and an approximate second-order Killing tensor that leads to a conserved Carter constant. These conditions restrict the set of arbitrary functions that parameterize the metric deformations to have a certain functional form. It is these conditions that select the appropriate ``bumps'' instead of the imposition of the Einstein equations. 

The restrictions imposed above are not strictly necessary, as there is no guarantee that BHs in alternative theories of gravity will continue to have three conserved quantities. However, all BH solutions in alternative theories of gravity known to date and that are not pathological (i.e.~they are stationary, axisymmetric, asymptotically flat, and contain no spacetime regions with closed time-like curves outside the horizon) possess three constants of the motion~\cite{Brown:2006pj}. Examples include the slowly-rotating solution found in dynamical CS gravity~\cite{Yunes:2009hc} and the spherically symmetric solution found in dynamical quadratic gravity~\cite{Yunes:2011we}. 

Much effort has gone into finding spacetimes with an {\emph{exact}} second-order (or higher-order) generalized ``Carter constant," thus allowing for the separation of the equations of motion~\cite{Brink:2008xx,Brink:2008xy,Brink:2009mq,Brink:2009mt,Brink:2009rf}. That work demonstrates that, for a broad class of spacetimes, such separation can be done provided that the Carter constant is {\emph{quartic}} in the orbit's 4-momentum (i.e., the constant is $C = \xi_{\alpha\beta\gamma\delta} p^\alpha p^\beta p^\gamma p^\delta$, where $\xi_{\alpha\beta\gamma\delta}$ is a 4th-rank Killing tensor and $p^{\alpha}$ is the 4-momentum). In this analysis, we show that one can in fact find an approximate Carter constant that is {\emph{quadratic}} in the 4-momentum, $C = \xi_{\alpha\beta} p^\alpha p^\beta$, for many relevant spacetimes, provided they differ from Kerr only {\it perturbatively}.  

Finally, we gain insight on the different proposed parameterizations by studying certain key limits. First, we show that the BK and DK deformations are related to each other by a gauge transformation. Second, we show that both the BK and DK metrics can be exactly mapped to specific non-GR BH metrics, the dynamical CS gravity one and the dynamical quadratic one. Third, we study the structure that the deformations must take when only deforming frame-dragging or the location of the event horizon. Fourth, we separate the geodesic equations into first-order form in both the BK and DK metrics, and calculate the modified Kepler law. This modification corrects the dissipative dynamics through the conversion of radial quantities to frequency space. 

The parameterized metrics (BK and DK) proposed in this paper lay the foundations for a systematic construction and study of ppE EMRI waveforms. With these metrics and their associated separated equations of motion, one can now study modified SCO trajectories and see how these impact the GW observable. Numerical implementation and a detailed data analysis study will be presented in a forthcoming publication. 

This paper is organized as follows. 
Sec~\ref{sec:BHs-GR} reviews BHs in GR and derives the first-order form of the geodesic equations.  
Sec.~\ref{sec:Standard-Bumps} introduces the BK formalism, which generalizes the bumpy BH formalism.
Sec.~\ref{sec:General-Kerr} presents the DK parameterized metric.
Sec.~\ref{sec:Comp} compares the parameterizations to each other and to alternative theory predictions.
Sec.~\ref{sec:Con} discusses some general properties of the parameterizations and points to future work.

In this paper we primarily follow the notation of Misner, Thorne and Wheeler~\cite{Misner:1973cw}. Greek letters stand for spacetime indices, while Latin letters in the middle of the alphabet $(i,j,k,\ldots)$ stand for spatial indices only. Parenthesis and brackets in index lists stand for symmetrization and antisymmetrization respectively, i.e.~$A_{(ab)} = (A_{ab} + A_{ba})/2$ and $A_{[ab]} = (A_{ab} - A_{ba})/2$. Background quantities are denoted with an overhead bar, such that the full metric $g_{\mu \nu} = \bar{g}_{\mu \nu} + \epsilon h_{\mu \nu}$ can be decomposed into a background $\bar{g}_{\mu \nu}$ plus a small deformation $h_{\mu \nu}$, where $\epsilon \ll 1$ is a book-keeping parameter. The background spacetime is always taken to be the Kerr metric. We use geometric units with $G = c = 1$. 

\section{Black Holes and Test-Particle Motion in General Relativity}
\label{sec:BHs-GR}

In this section, we review some basic material on the Kerr BH solution in GR 
and the motion of test-particles (see e.g.~\cite{Misner:1973cw} for more details). 
The most general stationary, axisymmetric BH solution in GR is the Kerr metric,
which in Boyer-Lindquist coordinates $(t,r,\theta,\phi)$ takes the form
\ba
\label{sol:metric_elements}
\bar{g}^{}_{tt} &=& - \left(1 - \frac{2 M r}{\rho^{2}} \right) \,, 
\qquad
\bar{g}^{}_{t\phi} = - \frac{2 M a r}{\rho^{2}} \sin^{2}{\theta}\,, 
\nonumber \\
\bar{g}^{}_{rr} &=& \frac{\rho^{2}}{\Delta}\,,
\qquad
\bar{g}^{}_{\theta \theta} = \rho^{2}\,,
\qquad
\bar{g}^{}_{\phi \phi} = \frac{\Sigma}{\rho^{2}} \sin^{2}{\theta}\,,
\ea
where we have introduced the quantities
\ba
\rho^{2} &\equiv& r^{2} + a^{2} \cos^{2}{\theta}\,, 
\\
\Delta &\equiv& r^{2} f + a^{2}\,, 
\qquad  
f \equiv 1 - \frac{2 M}{r}\,,
\\ 
\Sigma &\equiv& (r^{2} + a^{2})^{2} - a^{2} \Delta \sin^{2}{\theta}\,. 
\ea
The overhead bars are to remind us that the Kerr metric will be taken as our background spacetime. This background describes a BH with mass $M$ and spin angular momentum $S^{i} = (0,0,M a)$, where $a$ is the Kerr spin parameter with units of length. 

\subsection{Equations of Motion}
\label{subsec:bggeom-eom}

With this background, we can now compute the equations of motion for a test particle of mass $\mu$ moving on a background worldline $\bar{x}^{\mu} = \bar{z}^{\mu}(\lambda)$, where $\lambda$ is an affine parameter. One way to derive these equations is via the action for a non-spinning test-particle (see e.g.~\cite{Poisson:2004lr})
\be
S^{}_{\MAT} = - \mu \int^{}_{\gamma} d\lambda\; \sqrt{-\bar{g}^{}_{\alpha \beta}(\bar{z}) 
\dot{\bar{z}}^{\alpha} \dot{\bar{z}}^{\beta}} \,,
\ee
where $\dot{\bar{z}}^{\mu} = d \bar{z}^{\mu}/d\lambda$ is the tangent to $\bar{z}^{\mu}$ and $\bar{g}$ is the determinant of the background metric. 

The contribution of this action to the field equations can be obtained by varying it with respect to the metric tensor. Doing so, we obtain the stress-energy for the test-particle:
\be
T^{\alpha\beta}_{\MAT}(\bar{x}^{\mu}) = \mu\int \frac{d\tau}{\sqrt{-\bar{g}}}\bar{u}^\alpha \bar{u}^\beta
\updelta^{(4)}[\bar{\mb{x}}-\bar{\mb{z}}(\tau)] \,, 
\label{tmunu}
\ee
where $\tau$ is proper time, related to $\lambda$ via $d\tau =  
d\lambda\; \sqrt{-\bar{g}^{}_{\alpha \beta}(\bar{z}) \dot{\bar{z}}^{\alpha} \dot{\bar{z}}^{\beta}}$, 
$\bar{u}^{\mu} = d\bar{z}^{\mu}/d\tau$ is the particle's 4-velocity, normalized via 
$\bar{g}^{}_{\mu\nu}\bar{u}^{\mu}\bar{u}^{\nu}=-1$ and $\updelta^{(4)}$ is the four-dimensional
Dirac density, defined via $\int d^4x\sqrt{-\bar{g}}\,\updelta^{(4)}(\mb{x})=1$. 

The divergence of this stress-energy tensor must vanish in GR, which implies that test-particles follow geodesics:
\be
\frac{D}{d\tau}\frac{d\bar{z}^{\alpha}}{d\tau} = 0 \,,
\label{2nd-O-geodesics}
\ee
where $D/d\tau$ is a covariant derivative. The divergence of the stress-energy tensor vanishes due to local energy-momentum conservation and the equivalence principle~\cite{Misner:1973cw}. We have assumed here that the particle is non-spinning, so that geodesic motion is simply described by Eq.~\eqref{2nd-O-geodesics}; otherwise other terms would arise in the action that would lead to spin-dependent modifications.

\subsection{Constants of the Motion and Separability}

The geodesic equations as written in Eq.~\eqref{2nd-O-geodesics} are of second-order form, but they can be simplified to first-order form if there exist at least three constants of the motion plus a normalization condition on the four-momentum. If the metric is sufficiently symmetric, then Killing vectors $\xi^{\alpha}$ and Killing tensors $\xi^{\alpha \beta}$ exist that satisfy the vectorial and tensorial Killing equations
\ba
\nabla_{(\alpha} \xi_{\beta)} &=& 0 \,,
\\
\nabla_{(\alpha} \xi_{\beta \delta)} &=& 0 \,.
\label{Killing-Eq}
\ea

The Kerr metric possesses two Killing vectors and one Killing tensor. The vectors, $\bar{t}^{\alpha} =  [1,0,0,0]$ and $\bar{\phi}^{\alpha} = [0,0,0,1]$, are associated with the stationarity and axisymmetry of the metric respectively. In addition, the Kerr metric also possesses the Killing tensor:
\be
\bar{\xi}_{\alpha \beta} =  \Delta\, \bar{k}_{(\alpha} \bar{l}_{\beta)} + r^{2}\, \bar{g}^{}_{\alpha \beta}\,,
\label{KT-def}
\ee
where $\bar{k}^{\alpha}$ and $\bar{l}^{\alpha}$ are principal null directions:  
\ba
\bar{k}^{\alpha} &=& \left[\frac{r^{2}+a^{2}}{\Delta},1,0,\frac{a}{\Delta}\right]\,,
\nonumber \\
\bar{l}^{\alpha} &=& \left[\frac{r^{2}+a^{2}}{\Delta},-1,0,\frac{a}{\Delta}\right]\,.
\label{KPV}
\ea

The contraction of these Killing vectors and tensors allows us to define constants of the motion, namely the energy $E$, z-component of the angular momentum $L$, and the Carter constant $C$:
\ba
\label{E-def}
E &\equiv& - t^{\alpha} u_{\alpha}\,,
\\
\label{L-def}
L &\equiv& \phi^{\alpha} u_{\alpha}\,
\\
\label{C-def}
C &\equiv& \xi_{\alpha \beta} u^{\alpha} u^{\beta}\,.
\ea
It is conventional to define another version of the Carter constant, namely 
\be
Q \equiv C - \left(L - a E \right)^{2}\,.
\ee

With these three constants of the motion [either $(E,L,C)$ or $(E,L,Q)$] and the momentum condition $p^{\alpha} p_{\alpha} = - \mu^{2}$, or simply $u_{\alpha} u^{\alpha} = -1$, we can separate the geodesic equations and write them in first-order form:
\ba
\label{tdot-GR}
\rho^{2} \dot{\bar{t}} &=& \left[-a \left(a E \sin^{2}{\theta} - L \right) + \left(r^{2} 
+ a^{2} \right) \frac{P}{\Delta} \right]\,,  
\\
\label{phidot-GR}
\rho^{2} \dot{\bar{\phi}} &=& \left[- \left(a E - \frac{L}{\sin^{2}{\theta}} \right) 
+ \frac{a P}{\Delta} \right]\,,
\\
\label{rdot-GR}
\rho^{4}\, \dot{\bar{r}}^{2} &=&  \left[ \left( r^{2} + a^{2} \right) E - a L \right]^{2} 
\nonumber \\
&-&  \Delta\; \left[Q + \left(a E - L \right)^{2} + r^{2} \right]\,, 
\\
\label{thetadot-GR}
\rho^{4}\, \dot{\bar{\theta}}^{2} &=& Q - \cot^{2}{\theta} L^{2} 
- a^{2} \cos^{2}{\theta} \left(1 - E^{2} \right)\,,
\ea
where overhead dots stand for differentiation with respect to proper time and $P \equiv E (r^2+a^2)-aL$. Notice that not only have the equations been written in first-order form, but they have also been separated. 

Once the geodesic equations have been derived, one can compute Kepler's third law, relating radial separations to orbital frequencies for an object moving in an equatorial, circular orbit. For such an orbit, the equatorial, circular orbit conditions $\dot{\bar{r}} = 0$ and $d\dot{\bar{r}}/dr = 0$ lead to~\cite{Bardeen:1972fi}
\be
|\bar{\Omega}| \equiv |\dot{\bar{\phi}}/\dot{\bar{t}}| = \frac{M^{1/2}}{r^{3/2} + a M^{1/2}}\,.
\label{Kepler-Kerr}
\ee
In the far field limit, $M/r \ll 1$, this equation reduces to the familiar Kepler law: $\Omega \sim (M/r^{3})^{1/2}$.

\section{Black Holes and Test-Particle Motion in Alternative Theories: Generalized Bumpy Formalism} 
\label{sec:Standard-Bumps}

We here introduce the standard bumpy BH formalism~\cite{Collins:2004ex,2010PhRvD..81b4030V,2010arXiv1008.1278V} and generalize it to the BK framework. The bumps are constrained by requiring that an approximate, second-order Killing tensor exists. We then rewrite the geodesic equations in first-order form. 

\subsection{From Standard to Generalized Bumpy Black Holes}

Let us first review the basic concepts associated with the bumpy BH formalism~\cite{Collins:2004ex,2010PhRvD..81b4030V,2010arXiv1008.1278V}. This framework was initially introduced by Collins and Hughes~\cite{Collins:2004ex} to model deformations of the Schwarzschild metric and expanded by Vigeland and Hughes~\cite{2010PhRvD..81b4030V} to describe deformations on a Kerr background. In its standard formulation, generalized to spinning BHs~\cite{2010PhRvD..81b4030V}, the metric is {\emph{perturbatively}} expanded via   
\be
g_{\mu \nu} = \bar{g}_{\mu \nu} + \epsilon \; h_{\mu \nu}^{\SBK}\,,
\label{full-g}
\ee
where $\epsilon$ is a book-keeping parameter that reminds us that $|h_{\mu \nu}^{\SBK}|/|g_{\mu \nu}| \ll 1$. The background metric $\bar{g}_{\mu \nu}$ is the Kerr solution of Eq.~\eqref{sol:metric_elements}, while in the standard bumpy formalism, the metric functions $h_{\mu \nu}^{\SBK}$ are parameterized via 
\begin{widetext}
\ba
h_{tt}^{\SBK} &=& - 2 \psi^{\SBK}_{1} \left(1 + \frac{2 M r}{\rho^{2}} \right) - \frac{4 a M r}{\rho^{4}} \sigma^{\SBK}_{1}\,,
\qquad
h^{\SBK}_{tr} = - \gamma^{\SBK}_{1} \frac{2 a^{2} M r \sin^{2}{\theta}}{\Delta \rho^{2}}\,,
\nonumber \\
h^{\SBK}_{t \phi} &=& - \left(2 \psi^{\SBK}_{1} - \gamma^{\SBK}_{1}\right) \frac{2 a M r \sin^{2}{\theta}}{\rho^{2}} 
- 2 \sigma^{\SBK}_{1} \left[ \frac{\left(r^{2} + a^{2}\right) 2 M r }{\rho^{4}} + \frac{\Delta}{\rho^{2}} - \frac{\Delta}{2 \rho^{2} - 4 M r} \right]\,,
\nonumber \\
h^{\SBK}_{rr} &=& 2 \left(\gamma^{\SBK}_{1} - \psi^{\SBK}_{1} \right) \frac{\rho^{2}}{\Delta}\,,
\qquad
h^{\SBK}_{r \phi} =  \gamma^{\SBK}_{1} \left[1 + \frac{2 M r \left(r^{2} + a^{2}\right)}{\Delta \rho^{2}} \right] a \sin^{2}{\theta}\,,
\qquad
h^{\SBK}_{\theta \theta} = 2 \left(\gamma^{\SBK}_{1} - \psi^{\SBK}_{1} \right) \rho^{2}\,,
\nonumber \\
h^{\SBK}_{\phi \phi} &=& \left[\left(\gamma^{\SBK}_{1} - \psi^{\SBK}_{1} \right) \frac{8 a^{2} M^{2} r^{2} \sin^{2}{\theta}}{\rho^{2} - 2 M r} - 2 \psi^{\SBK}_{1} \frac{\rho^{4} \Delta}{\rho^{2} - 2 M r} 
+ \frac{4 a M r}{\rho^{2} - 2 M r} \sigma^{\SBK}_{1} \left(\Delta + \frac{2 M r \left(r^{2} + a^{2} \right)}{\rho^{2}} \right) \right] \frac{\sin^{2}{\theta}}{\rho^{2}}\,,
\label{Bumpy-hab}
\ea
\end{widetext}
where $(\psi^{\SBK}_{1},\gamma^{\SBK}_{1},\sigma^{\SBK}_{1})$ are functions of $(r,\theta)$ only that represent the ``standard bumps.''

A few comments are due at this point. First, notice that this metric contains only three arbitrary functions $(\psi^{\SBK}_{1},\gamma^{\SBK}_{1},\sigma^{\SBK}_{1})$, instead of four, as one would expect from the most general stationary and axisymmetric metric. This is because of the specific way Eq.~\eqref{Bumpy-hab} is derived (see below), which assumes a Ricci-flat metric in the $a=0$ limit. Second, note that many metric components are non-vanishing: apart from the usual $(t,t)$, $(t,\phi)$, $(r,r)$, $(\theta,\theta)$ and $(\phi,\phi)$ metric components that are non-zero in the Kerr metric, Eq.~\eqref{Bumpy-hab} also has non-vanishing $(t,r)$ and $(r,\phi)$ components.

The derivation of this metric is as follows. One begins with the most general stationary and spherically symmetric metric in Lewis-Pappapetrou form that is Ricci-flat. One then perturbs the two arbitrary functions in this metric and maps it to Schwarschild coordinates. Through the Newman-Janis procedure, one then performs a complex rotation of the tetrad associated with this metric to transform it into a deformed Kerr spacetime. At no stage in this procedure is one guaranteed that the resulting spacetime will still possess a Carter constant or that it will remain vacuum. In fact, the metric constructed from Eq.~\eqref{Bumpy-hab} does not have a Carter constant, nor does it satisfy the vacuum Einstein equations, leading to a non-zero effective stress-energy tensor.

Such features make the standard bumpy formalism not ideal for null-tests of GR. One would prefer to have a framework that is generic enough to allow for non-GR tests while still possessing a smooth GR limit, such that as the deformations are taken to zero, one recovers exactly the Kerr metric. This limit implies that the deformed metric will retain many of the nice properties of the Kerr background, such as the existence of an event horizon, an ergosphere, and a Carter constant. 

This can be achieved by generalizing the bumpy BH formalism through the relaxation of the initial assumption that the metric be Ricci-flat prior to the Newman-Janis procedure. Inspired by~\cite{Collins:2004ex,2010PhRvD..81b4030V,2010arXiv1008.1278V}, we therefore promote the non-vanishing components of the metric perturbation, i.e.,~$(h^{\BK}_{tt}, h^{\BK}_{tr}, h^{\BK}_{t\phi},h^{\BK}_{rr},h^{\BK}_{r \phi},h^{\BK}_{\theta \theta},h^{\BK}_{\phi \phi})$, to arbitrary functions of $(r,\theta)$. These functions are restricted only by requiring that an approximate second-order Killing tensor still exists. 

This restriction is not strictly necessary, but it is appealing on several fronts. First, the few analytic non-GR BH solutions that are known happen to have a Carter constant, at least perturbatively as an expansion in the non-Kerr deviation. Second, a Carter constant allows for the separation of the equations of motion into first-order form, which then renders the system easily integrable with already developed numerical techniques. 

\subsection{Existence Conditions for the Carter Constant}
\label{subsec:C-const}

Let us now investigate what conditions must be enforced on the metric perturbation $h^{\BK}_{\alpha \beta}$ so that a Killing tensor $\xi_{\alpha \beta}$ and its associated Carter constant exist, at least perturbatively to ${\cal{O}}(\epsilon)$.  Killing's equations can be used to infer some fairly general properties that such a tensor must have. First, this tensor must be non-vanishing in the same components as the metric perturbation. For the generalized bumpy metric, this means that $(\xi_{t\theta},\xi_{r\theta},\xi_{\theta \phi})$ must all vanish. Furthermore, if we expand the Killing tensor as
\be
\xi_{\alpha \beta} = \bar{\xi}_{\alpha \beta} + \epsilon \; \delta \xi_{\alpha \beta}\,,
\ee
we then find that $\delta \xi_{\alpha \beta}$ must have the same parity as $\bar{\xi}_{\alpha \beta}$ for the full Killing tensor to have a definite parity. For the generalized bumpy metric, this means that $(\xi_{tt},\xi_{rr},\xi_{\theta\theta},\xi_{\phi\phi},\xi_{t\phi})$ must be even under reflection: $\theta \to \theta - \pi$. 

These conditions imply that of the $10$ independent degrees of freedom in $\delta \xi_{\alpha \beta}$, only 7 are truly necessary. With this in mind, we parameterize the Killing tensor as in Eq.~\eqref{KT-def}, namely
\be
\label{KT-parameterization}
\xi_{\alpha \beta} =  \Delta\, k_{(\alpha} l_{\beta)} + r^{2}\, {g}^{}_{\alpha \beta}\,.
\ee
Notice that $\xi_{\alpha \beta}$ depends here on the full metric $g_{\alpha \beta}$ and on
null vectors $k^{\alpha}$ and $l^{\alpha}$ that are not required to be the Kerr ones or the principal congruences of the full spacetime: $k^{\alpha} \neq \bar{k}^{\alpha}$ and $l^{\alpha} \neq \bar{l}^{\alpha}$. We decompose these vectors via
\ba
\label{null-vec-decomp}
k^{\alpha} = \bar{k}^{\alpha} + \epsilon \; \delta k^{\alpha}\,,
\qquad
l^{\alpha} = \bar{l}^{\alpha} + \epsilon \; \delta l^{\alpha}\,,
\ea
where $\bar{k}^{\alpha}$ and $\bar{l}^{\alpha}$ are given in Eq.~\eqref{KPV}, and the 
tensor $\xi_{\alpha \beta}$ into
\be
\xi_{\alpha \beta} = \bar{\xi}_{\alpha \beta} + \epsilon \; \delta \xi_{\alpha \beta}\,,
\ee
where $\bar{\xi}_{\alpha \beta}$ is given in Eq.~\eqref{KT-def}, while
\ba
\delta \xi_{\alpha \beta} &\equiv& \Delta \left[ \delta k_{(\alpha} l_{\beta)} + \delta l_{(\alpha} k_{\beta)} + 2 h^{\BK}_{\delta (\alpha} \bar{k}_{\beta)} \bar{l}^{\delta} \right]
\nonumber \\
&& + 3 r^{2} h^{\BK}_{\alpha \beta} \,.
\ea
All lowering and raising of indices is carried out with the background metric. Although this particular ansatz only allows $6$ independent degrees of freedom (as the vectors are assumed to be null), we will see it suffices to find a Carter constant. 

With this ansatz, the tensor Killing equation [Eq.~\eqref{Killing-Eq}] becomes  
\ba
\partial_{(\mu} \delta \xi_{\alpha \beta)} - 2 \, \bar{\Gamma}^{\delta}_{(\mu \alpha} \, \delta\xi_{\beta) \delta} =  2 \, \delta \Gamma^{\delta}_{(\mu \alpha} \, \bar{\xi}_{\beta) \delta}\,.
\label{red-KT-Eq}
\ea
The system of equations one must solve is truly formidable and in fact overconstrained. Equation~\eqref{red-KT-Eq} is a set of 20 partial differential equations, while the normalization conditions $l^{\alpha} l_{\alpha} = 0 = k^{\alpha} k_{\alpha}$ add two additional algebraic equations. This means there are a total of 13 degrees of freedom (7 in $h^{\BK}_{\alpha \beta}$ and $6$ in $\delta l^{\alpha}$ and $\delta k^{\alpha}$ after imposing the normalization condition) but 20 partial differential equations to solve.

In spite of these difficulties, we have solved this system of equations with Maple and the GrTensorII package~\cite{grtensor} and found that the perturbation to the null vectors must satisfy
\ba
\delta k_{\BK}^{\alpha} &=& \left[\frac{r^{2}+a^{2}}{\Delta} \delta k_{\BK}^r + \delta_1, \delta k_{\BK}^r , 0 , \frac{a}{\Delta} \delta k_{\BK}^r + \delta_2\right]\,, \nonumber
\\
\delta l_{\BK}^{\alpha} &=& \left[\frac{r^{2}+a^{2}}{\Delta} \delta k_{\BK}^r + \delta_3, \delta k_{\BK}^r + \delta_4 , 0 , \frac{a}{\Delta} \delta k_{\BK}^r + \delta_5\right]\,, \nonumber \\
\ea
where $\delta_{i}\equiv\delta_{i}(r)$ are arbitrary functions of $r$, generated upon solving the differential system. These functions are fully determined by the metric perturbation, which is given by
\begin{widetext}
\ba
\label{BK-metric-exp}
h^{\BK}_{tt} &=& -\frac{a}{2M}\frac{\rho^4 \Delta}{P^{\BK}_1}\frac{\partial h^{\BK}_{t\phi}}{\partial r} - \frac{a}{M} \frac{P^{\BK}_2}{P^{\BK}_1} h^{\BK}_{t\phi} - \frac{a^2\sin^2\theta}{4M}\frac{(\rho^2-4Mr)\Delta^2}{P^{\BK}_1}
\left(\frac{\partial \delta_1}{\partial r} + \frac{\partial \delta_3}{\partial r}\right) - \frac{a}{4M}\Delta^2\sin^2\theta\frac{P^{\BK}_3}{P^{\BK}_1}\left(\frac{\partial \delta_2}{\partial r} +  \frac{\partial \delta_5}{\partial r} \right) \nonumber \\
&& + \frac{\Delta}{\rho^2}\frac{P^{\BK}_4}{P^{\BK}_1}(\delta_1+\delta_3) - \frac{a}{2M}\frac{\Delta}{\rho^2}\frac{P^{\BK}_5}{P^{\BK}_1}(\delta_2+\delta_5) + \frac{\Delta}{\rho^2}\frac{(r^2+a^2)\bar{\rho}^{2}}{P^{\BK}_1}\Theta_1 \,,
\nonumber
\ea
\ba
h^{\BK}_{tr} &=& \frac{1}{2}\left(1-\frac{2Mr}{\rho^2}\right)(\delta_1-\delta_3) + \frac{aMr\sin^2\theta}{\rho^2}(\delta_2-\delta_5) - \frac{1}{2}\delta_4 \,,
\qquad
h^{\BK}_{rr} = -\frac{\Theta_1}{\Delta} + \frac{\rho^2}{\Delta} \delta_4 \,,
\nonumber \\
h^{\BK}_{r\phi} &=& \frac{aMr\sin^2\theta}{\rho^2}(\delta_1-\delta_3) - \frac{\sin^2\theta}{2\rho^2} (P^{\BK}_3-2a^2Mr\sin^2\theta)(\delta_2-\delta_5) + \frac{a}{2}\sin^2\theta \delta_4 \,,
\nonumber \\
h^{\BK}_{\theta\theta} &=& -\Theta_1 + \rho^{2} \Theta_2 \,,
\nonumber \\
h^{\BK}_{\phi\phi} &=& -\frac{(r^2+a^2)^2}{a^2}h^{\BK}_{tt} - \frac{2(r^2+a^2)}{a} h^{\BK}_{t\phi} + \frac{\Delta}{a^2} \Theta_1 + \frac{\Delta^2}{a^2}(\delta_1+\delta_3) - \frac{\Delta^2\sin^2\theta }{a}(\delta_2+\delta_5) \,,
\nonumber \\
\frac{\partial^{2} h^{\BK}_{t\phi}}{\partial r^{2}} &=& 
\left( \frac{\partial^{2} \delta_{2}}{\partial r^{2}}  + \frac{\partial^{2} \delta_{5}}{\partial r^{2}} \right) P^{\BK}_{6} +  \left( \frac{\partial^{2} \delta_{1}}{\partial r^{2}} + \frac{\partial^{2} \delta_{3}}{\partial r^{2}}\right) P^{\BK}_{7} + \left( \frac{\partial \delta_{2}}{\partial r}  + \frac{\partial \delta_{5}}{\partial r} \right) \frac{P^{\BK}_{8,n}}{P^{\BK}_{8,d}} + 
\left( \frac{\partial \delta_{1}}{\partial r} + \frac{\partial \delta_{3}}{\partial r}\right) \frac{P^{\BK}_{9,n}}{P^{\BK}_{9,d}} 
\nonumber \\
&& + \left( \delta_{2} + \delta_{5} \right) \frac{P^{\BK}_{10,n}}{P^{\BK}_{10,d}} + 
\left( \delta_{1} + \delta_{3} \right) \frac{P^{\BK}_{11,n}}{P^{\BK}_{11,d}} 
+ \frac{\partial h^{\BK}_{t\phi}}{\partial r} \frac{P^{\BK}_{12,n}}{P^{\BK}_{12,d}} 
+ h^{\BK}_{t\phi} \frac{P^{\BK}_{13,n}}{P^{\BK}_{13,d}} + \Theta_{1} P^{\BK}_{14}\,,
\ea
\end{widetext}
where $\bar{\rho}^{2}\equiv r^{2} - a^{2} \cos^{2}{\theta}$, $\Theta_{1,2} \equiv \Theta_{1,2}(\theta)$ are functions of $\theta$ and $P^{\BK}_i$ are polynomials in $r$ and $\cos\theta$, which are given explicitly in Appendix~\ref{app:pols}. Notice that $\delta k_{\BK}^{r}$ does not appear in the metric perturbation at all, so we are free to set it to zero, i.e.~$\delta k_{\BK}^r=0$. 

With this at hand, given some metric perturbation $h^{\BK}_{\alpha \beta}$ that satisfies Eq.~\eqref{BK-metric-exp}, one can construct the arbitrary functions $\delta_{i}$ and $\Theta_{i}$ and thus build the null directions of the perturbed spacetime, such that a Killing tensor and a Carter constant exist perturbatively to ${\cal{O}}(\epsilon)$. We have verified that the conditions described above satisfy the generic Killing tensor properties, described at the beginning of this subsection. 

\subsubsection{Non-Rotating Limit}
\label{BK-non-rot}

Let us now investigate the non-rotating limit $a \to 0$. To do so, we expand all arbitrary functions via
\be
{\cal{F}}_{i} = {\cal{F}}_{i,0} + a {\cal{F}}_{i,1} + {\cal{O}}(a^{2})\,,
\label{low-a-exp-arb}
\ee
where ${\cal{F}}_{i}$ is any of $\delta_{i}(r)$ or $\Theta_{i}(\theta)$. Let us also expand the metric components in a similar fashion, i.e.~
\be
h^{\BK}_{\alpha \beta}(r,\theta) = h^{\BK}_{\alpha \beta,0}(r,\theta) + a \, h^{\BK}_{\alpha \beta,1}(r,\theta) + {\cal{O}}(a^{2})\,.
\label{low-a-exp-g}
\ee

The following choices
\ba
\Theta_{1,0} &=& 0 = \Theta_{2,n}\,,
\qquad
\delta_{1,0} =  \frac{h_{rr,0}^{\BK}}{2} + \frac{h_{tt,0}^{\BK}}{2 f^{2}}\,, 
\nonumber \\
\delta_{2,0} &=& 0 = \delta_{5,0}\,,
\quad
\delta_{2,1} = \delta_{5,1} = -\frac{r-4 M}{2 r^{3} f^{2}} h^{\BK}_{tt,0}\,,
 \nonumber \\
 \delta_{4,0} &=& f \, h^{\BK}_{rr,0} \,, 
\qquad
 \delta_{3,0} = - \frac{h^{\BK}_{rr,0}}{2} + \frac{h^{\BK}_{tt,0}}{2 f^{2}}\,,
\ea 
where we recall that $f \equiv 1 - 2 M/r$, force the metric components to take the form
\ba
h^{\BK}_{tt} &=& h^{\BK}_{tt,0}(r)\,,
\qquad 
h^{\BK}_{rr} = h^{\BK}_{rr,0}(r)\,,
\ea
where we have set all integration constants to zero. All other components of the metric vanish to this order. To obtain this result, it is crucial to use the slow-rotation expansion postulated above, as some of the conditions in Eq.~\eqref{BK-metric-exp} have seemingly divergent pieces. We see then that in the non-rotating limit, one can always choose free functions $\delta_{i}$ and $\Theta_{i}$ such that the only two independent metric perturbations are $h_{tt}^{\BK}$ and $h_{rr}^{\BK}$.

The Killing tensor is then given by Eq.~\eqref{KT-parameterization}, with the perturbed null vector components
\ba
\delta k_{\BK}^{\alpha} &=& \left[ \frac{h^{\BK}_{rr,0}}{2} + \frac{h^{\BK}_{tt,0}}{2 f^{2}}, 0, 0, {\cal{O}}(a)\right]\,,
\\ \nonumber
\delta l_{\BK}^{\alpha} &=& \left[ - \frac{h^{\BK}_{rr,0}}{2} + \frac{h^{\BK}_{tt,0}}{2 f^{2}}, 
f \, h^{\BK}_{rr,0}, 0 , {\cal{O}}(a) \right]\,,
\ea
where we have set $\delta k_{\BK}^{r} = 0$.  Composing the Killing tensor, we find that all ${\cal{O}}(\epsilon)$ terms vanish in the $a \to 0$ limit. This result makes perfect sense, considering that the Schwarzschild Killing tensor, $\bar{\xi}_{\alpha \beta}= r^{2} \Omega_{\alpha \beta}$ with $\Omega_{\alpha \beta} \equiv {\rm{diag}}(0,0,r^{2},r^{2} \sin^{2}{\theta})$, is also a Killing tensor for the most generic static and spherically symmetric spacetime with arbitrary $(t,t)$ and $(r,r)$ components. 

\subsection{Equations of Motion and Geodesics in First-Order Form}

The equations of motion for test-particles in alternative theories of gravity need not be geodesic. But as we review in Appendix~\ref{alt-theories-EOM}, they can be approximated as geodesics in the test-particle limit for a wide class of alternative theories, provided we neglect the spin of the small body. Here, we restrict attention to theories where the modified equations of motion remain geodesic, but of the generalized bumpy Kerr background constructed above.

The geodesic equations can be rewritten in first-order form provided that three constants of the motion exist. In Sec.~\ref{subsec:C-const}, we showed that provided the generalized bumpy functions satisfy certain conditions, then an approximate, second-order Killing tensor and a conserved Carter constant exist. Furthermore, it is easy to see that the metric in Eqs.~\eqref{full-g} and~\eqref{Bumpy-hab} remains stationary and axisymmetric, since $(h^{\BK}_{tt}, h^{\BK}_{tr}, h^{\BK}_{t\phi},h^{\BK}_{rr},h^{\BK}_{r \phi},h^{\BK}_{\theta \theta},h^{\BK}_{\phi \phi})$ are functions of only $(r,\theta)$. 

The constants of the motion then become
\ba
\label{EqE}
E &=& \bar{E} +\epsilon \left( h_{\mu \nu} t^{\mu} \bar{u}^{\nu} + \bar{g}_{\mu \nu} t^{\mu} \delta u^{\nu} \right)\,,
\\
L &=& \bar{L} +\epsilon \left( h_{\mu \nu} \phi^{\mu} \bar{u}^{\nu} + \bar{g}_{\mu \nu} \phi^{\mu} \delta u^{\nu} \right)\,,
\\
C &=& \bar{C} +\epsilon \left( \delta \xi_{\mu \nu} \bar{u}^{\mu} \bar{u}^{\nu} + 2 \bar{\xi}_{\mu \nu} \bar{u}^{(\mu} \delta u^{\nu)} \right)\,,
\label{EqC}
\ea
while the normalization condition becomes
\be
\label{norm-cond}
0 = h_{\mu \nu} \bar{u}^{\mu} \bar{u}^{\nu}
+ 2 \bar{g}_{\mu \nu} \bar{u}^{\mu} \delta u^{\nu}\,,
\ee
since by definition $-1 = \bar{g}_{\mu \nu} \bar{u}^{\mu} \bar{u}^{\nu}$. In these equations, $\bar{u}^{\mu} = [\dot{\bar{t}},\dot{\bar{r}},\dot{\bar\theta},\dot{\bar\phi}]$ is the unperturbed, Kerr four-velocity, while $\delta u^{\mu}$ is a perturbation. 

Equations~\eqref{EqE}-\eqref{norm-cond} can be decoupled once we make a choice for $(E,L,Q)$. This choice affects whether the turning points in the orbit are kept the same as in GR, or whether the constants of the motion are kept the same. We here choose the latter for simplicity by setting $E = \bar{E}$, $L = \bar{L}$ and $C = \bar{C}$, which then implies that $\delta u^{\mu}$ must be such that all terms in parenthesis in Eqs.~\eqref{EqE}-\eqref{EqC} vanish. Using this condition and Eq.~\eqref{norm-cond}, we then find that 
\ba
\rho^2\dot{t} &=& T_K(r,\theta) + \delta T(r,\theta) \,, \nonumber \\
\qquad \rho^4\dot{r}^2 &=& R_K(r) + \delta R(r,\theta) \,,\nonumber \\
\rho^2\dot{\phi} &=& \Phi_K(r,\theta) + \delta \Phi(r,\theta) \,, \nonumber \\
\qquad \rho^4\dot{\theta}^2 &=& \Theta_K(\theta) + \delta \Theta(r,\theta) \,,
\ea
where $(T_K, R_K, \Theta_K, \Phi_K)$ are given by the right-hand sides of Eqs. (\ref{tdot-GR}), (\ref{rdot-GR}), (\ref{thetadot-GR}), and (\ref{phidot-GR}) respectively. The perturbations to the potential functions $(\delta T, \delta R, \delta\Theta, \delta\Phi)$ are given by
\ba
\delta T(r,\theta) &=& \left[ \frac{(r^2+a^2)^2}{\Delta} - a^2\sin^2\theta \right] h_{t\alpha}\bar{u}^\alpha 
\nonumber \\
&+& \frac{2aMr}{\Delta} h_{\phi\alpha} \bar{u}^\alpha \,, \label{dT_potential} \\
\delta \Phi(r,\theta) &=& \frac{2aMr}{\Delta} h_{t\alpha} \bar{u}^\alpha - \frac{\rho^2-2Mr}{\Delta\sin^2\theta} h_{\phi\alpha} \bar{u}^\alpha \,, \label{dPhi_potential} \\
\delta R(r,\theta) &=& \Delta \left[A(r,\theta) \, r^2+B(r,\theta)\right] \,, \\
\delta \Theta(r,\theta) &=& A(r,\theta) \, a^2\cos^2\theta - B(r,\theta) \,,
\ea
where the functions $A(r,\theta)$ and $B(r,\theta)$ are proportional to the perturbation:
\ba
A(r,\theta) &=& 2 \left[ h_{\alpha t} \bar{\dot{t}} + h_{\alpha \phi} \bar{\dot{\phi}} \right]  \bar{u}^\alpha - h_{\alpha\beta} \bar{u}^\alpha \bar{u}^\beta \,, \\
B(r,\theta) &=& 2 \left[ \left( \bar{\xi}_{tt} \bar{\dot{t}} + \bar{\xi}_{t\phi} \bar{\dot{\phi}} \right) \delta\dot{t} + \left( \bar{\xi}_{t\phi} \bar{\dot{t}} + \bar{\xi}_{\phi\phi} \bar{\dot{\phi}} \right) \delta\dot{\phi}  \right] \nonumber \\
	&& + \delta\xi_{\alpha\beta} \bar{u}^\alpha \bar{u}^\beta \,.
\ea
As before, we have dropped the superscript $BK$ here. Interestingly, notice that the perturbation automatically couples the $(r,\theta)$ sector, so that the first-order equations are not necessarily separable. One can choose, however, the $\gamma_{i}$ functions in such a way so that the equations remain separable, as will be shown elsewhere.

The perturbations to the potential functions result in a modification of Kepler's law [Eq.\ (\ref{Kepler-Kerr})]:
\ba
|\Omega| &=& |\bar{\Omega}|  -  \frac{M^{1/2} \left(r^{1/2}(r-3M)+2aM^{1/2}\right)}{r^{5/4} \left(r^{3/2}+aM^{1/2}\right)^2} \delta T(r) 
\nonumber \\
&+& \frac{\left(r^{1/2}(r-3M)+2aM^{1/2}\right)^{1/2}}{r^{5/4} \left(r^{3/2}+aM^{1/2}\right)} \delta\Phi(r)\,,
\label{mod-Kepler-law}
\ea
where we recall that $|\bar{\Omega}|$ was already given in Eq.~\eqref{Kepler-Kerr}, $\delta T$ and $\delta \Phi$ are given by Eq.~(\ref{dT_potential}) and (\ref{dPhi_potential}), and we have used that $E = \bar{E}$ and $L = \bar{L}$. In the far field limit ($M/r \ll 1$), this becomes
\be
|\Omega| \sim |\bar{\Omega}| + \frac{1}{r^2} \delta\Phi \;.
\ee
where we have assumed that $\delta \Phi$ and $\delta T$ are both of order unity for simplicity. Given any particular metric perturbation, one can easily recalculate such a correction to Kepler's law from Eq.~\eqref{mod-Kepler-law}. Clearly, a modification of this type in the Kepler relation automatically modifies the dissipative dynamics when converting quantities that depend on the orbital frequency to radius and vice-versa. 

\section{Black Holes and Test-Particle Motion in Alternative Theories: Deformed Kerr Formalism}
\label{sec:General-Kerr}

In the previous section, we generalized the bumpy BH framework at the cost of introducing a large number of arbitrary functions, later constrained by the requirement of the existence of a Carter constant. In this section, we investigate a different parameterization (DK) that isolates the physically independent degrees of freedom from the start so as to minimize the introduction of arbitrary functions. 

\subsection{Deformed Kerr Geometry}

Let us first consider the most general spacetime metric that one can construct for a stationary and axisymmetric spacetime. In such a geometry, there will exist two Killing vectors, $t^{a}$ and $\phi^{a}$,  that represent invariance under a time and azimuthal coordinate transformation. Because these Killing vectors are independent, they will commute satisfying $t_{[a} \phi_{b} \nabla_{c} t_{d]} = 0 = t_{[a} \phi_{b} \nabla_{c} \phi_{d]}$. Let us further assume the integrability condition
\be
t^{a} R_{a}{}^{[b} t^{c} \phi^{d]} = 0 = \phi^{a} R_{a}{}^{[b} t^{c} \phi^{d]}\,.
\label{extra-cond}
\ee

The condition in Eq.~\eqref{extra-cond} guarantees that the $2$-planes orthogonal to the Killing vectors $t^{a}$ and $\phi^{a}$ are integrable. Generic  stationary and axisymmetric solutions to modified field equations do not need to satisfy Eq.~\eqref{extra-cond}. However, all known analytic solutions in GR and in alternative theories do happen to satisfy this condition. We will thus assume it holds here as well. 

Given these conditions, the most general stationary and axisymmetric line element can be written in Lewis-Papapetrou canonical form, using cylindrical coordinates $(t,\rho,\phi,z)$:
\be
ds^{2} = -V \left(dt - w d\phi\right)^{2} + V^{-1} \rho^{2} d\phi^{2} + \Omega^{2} \left( d\rho^{2} + \Lambda dz^{2} \right)\,,
\label{Lp-form}
\ee
where $(V,w,\Omega,\Lambda)$ are arbitrary functions of $(\rho,z)$. In GR, by imposing Ricci-flatness, one can eliminate one of these functions via a coordinate transformation, and we can thus set $\Lambda = 1$. The Einstein equations then further restrict the form of these arbitrary functions~\cite{Jones:2005hj}
\ba
\bar{V} &=& \frac{\left(r_{+} + r_{-}\right)^{2} - 4 M^{2} + \frac{a^{2}}{M^{2}-a^{2}} \left(r_{+} - r_{-}\right)^{2}}{\left(r_{+} + r_{-} + 2 M\right)^{2} + \frac{a^{2}}{M^{2}-a^{2}} \left(r_{+} - r_{-}\right)^{2}}\,,
\nonumber \\
\bar{w} &=& \frac{2 a M \left(M + \frac{r_{+} + r_{-}}{2} \right) \left(1 - \frac{\left(r_{+} - r_{-}\right)^{2}}{4 \left(M^{2} - a^{2}\right)} \right)}{\frac{1}{4} \left(r_{+} + r_{-} \right)^{2} - M^{2} + a^{2} \frac{\left(r_{+} - r_{-} \right)^{2}}{4 \left(M^{2} - a^{2} \right)}}\,,
\nonumber \\
\bar{\Lambda} &=& 1\,,
\nonumber \\
\bar{\Omega} &=& V \frac{\left(r_{+} + r_{-} \right)^{2} - 4 M^{2} + \frac{a^{2}}{M^{2}-a^{2}} \left(r_{+} - r_{-}\right)^{2}}{4 r_{+} r_{-}}\,,
\ea
where $r_{\pm} = \sqrt{\rho^{2} + \left(z \pm \sqrt{M^{2} - a^{2}}\right)^{2}}$. One can map from this coordinate system to Boyer-Lindquist coordinates $(t,r,\theta,\phi)$ via the transformation $\rho = \sqrt{\Delta} \sin{\theta}$ and $z = (r - M) \sin{\theta}$ to find the metric in Eq.~\eqref{sol:metric_elements}. 

We will not impose Ricci-flatness here, as we wish to model alternative theory deviations from Kerr. We thus work directly with Eq.~\eqref{Lp-form}, keeping $(V,w,\Omega,\Lambda)$ as arbitrary functions, but transforming this to Boyer-Lindiquist coordinates. Let us then re-parametrize $\Omega^{2} \equiv \gamma$, $\lambda \equiv \gamma \Lambda$, and $q \equiv V w$, and perturb the metric functions via 
\ba
V &=& \bar{V} + \delta V\,, 
\qquad
q = \bar{q} + \delta q\,,
\nonumber \\
\lambda &=& \bar{\lambda} + \delta \lambda\,,
\qquad
\gamma = \bar{\gamma} + \delta \gamma\,.
\ea
The metric then becomes 
\be
\label{full-g2}
g_{\mu \nu} = \bar{g}_{\mu \nu}+ h^{\DK}_{\mu \nu}\,,
\ee
where the perturbation is given by
\ba 
\label{DK-hab}
h_{tt}^{\DK} &=& - \delta V\,,
\qquad
h_{t \phi}^{\DK} = \delta q\,,
\nonumber \\
h^{\DK}_{rr} &=& \delta \lambda \cos^{2}{\theta} + \delta\gamma \frac{(r-M)^2\sin^2\theta}{\Delta} \,,
\nonumber \\
h^{\DK}_{r\theta} &=& \left(r-M\right) \cos{\theta} \sin{\theta} \left( \delta \gamma - \delta \lambda \right)\,,
\nonumber \\
h^{\DK}_{\theta \theta} &=& \delta \lambda \sin^{2}{\theta} \left(r-M\right)^{2} + \delta \gamma \cos^{2}{\theta} \Delta\,,
\nonumber \\
h^{\DK}_{\phi \phi} &=& \frac{\sin^2\theta}{\rho^{2}-2Mr} \left\{ 4aMr \, \delta q \right.
\nonumber \\
	&& \left. - \left[ \left(r^2+a^2\right)^2 - a^2\sin^2\theta \Delta \right] \, \delta V \right\} \,,
\ea
and all other components vanish. 

As is clear from the above expressions, the most general perturbation to a stationary, axisymmetric metric yields five non-vanishing metric components that depend still only on four arbitrary functions 
$(\delta V, \delta q, \delta\gamma, \delta \lambda)$ of two coordinates $(r,\theta)$. 
This is unlike the standard bumpy formalism that introduces six non-vanishing metric components that depend on four arbitrary functions (courtesy of the Newman-Janis algorithm applied to a non-Kerr metric). Furthermore, this analysis shows that two of the six arbitrary functions introduced in the generalized bumpy scheme of the previous section must not be independent, as argued earlier. 

Following the insight from the previous section, we will henceforth allow these $5$ metric components $(h_{tt}^{\DK},h_{rr}^{\DK},h_{r\theta}^{\DK},h_{\theta\theta}^{\DK},h_{\phi\phi}^{\DK})$ to be arbitrary functions of $(r,\theta)$, although technically there are only four independent degrees of freedom. We do this because it eases the analytic calculations to come when one investigates which conditions $h_{\alpha \beta}^{\DK}$ must satisfy for there to exist an approximate second-order Killing tensor. Moreover, it allows us to relax the undesirable requirement, implicit in Eq.~\eqref{DK-hab}, that when $\delta V \neq 0$, then both $h_{tt}$ and $h_{\phi \phi}$ must be non-zero in the $a \to 0$ limit.

\subsection{Existence Conditions for the Carter constant}
Let us now follow the same methodology of Sec.~\ref{subsec:C-const} to determine what conditions the arbitrary functions must satisfy in order for there to be a second-order Killing tensor. We begin by parameterizing the Killing tensor just as in Eq.~\eqref{KT-parameterization}, with the expansion of the null vectors as in Eq.~\eqref{null-vec-decomp} and the replacement of $BK \to DK$ everywhere. With this at hand, the Killing equation acquires the same structure as Eq.~\eqref{red-KT-Eq}. 

We have solved the Killing equations with Maple and the GrTensorII package~\cite{grtensor} to find that the null vectors must satisfy:
\ba
\delta k_{\DK}^{\alpha} &=& \left[ \frac{r^{2} + a^{2}}{\Delta} \left(\delta l_{\DK}^{r} + \gamma_{1}\right) + \gamma_{4},
\delta l_{\DK}^{r} + \gamma_{1},
-\frac{h^{\DK}_{r\theta}}{\rho^{2}} ,
\right. 
\nonumber \\
&&  \left.
 \frac{a}{\Delta} \left(\delta l_{\DK}^{r} + \gamma_{1} \right) + \gamma_{3}\right]\,,
\nonumber \\
\delta l_{\DK}^{\alpha} &=& \left[  - \frac{r^{2} + a^{2}}{\Delta} \delta l_{\DK}^{r} + \gamma_{4}, \delta l_{\DK}^{r}, 
\frac{h^{\DK}_{r\theta}}{\rho^{2}}, - \frac{a}{\Delta} \delta l_{\DK}^{r} + \gamma_{3} \right]\,,
\nonumber \\
\ea
where $\gamma_{i}\equiv\gamma_{i}(r)$ are arbitrary functions of radius, while $\delta l_{\DK}^{r}$ and $h_{r\theta}^{\DK}$ are arbitrary functions of both $(r,\theta)$. 
As before with $\delta k_{\BK}^{r}$, we find below that the function $\delta l_{\DK}^{r}$ does not enter the metric perturbation, so we can set it to zero. 

The functions $\gamma_i$ are completely determined by the metric perturbation:
\begin{widetext}
\ba
	h^{\DK}_{tt} &=& -\frac{a}{M} \frac{P^{\DK}_2}{P^{\DK}_1} h^{\DK}_{t\phi} - \frac{a}{2M} \frac{\rho^4 \Delta}{P^{\DK}_1} \frac{\partial h^{\DK}_{t\phi}}{\partial r} - \frac{2a^2r (r^2+a^2) \Delta \sin\theta \cos\theta}{\rho^2 P^{\DK}_1} h^{\DK}_{r\theta} + \frac{(r^2+a^2)\bar{\rho}^{2} \Delta}{\rho^2 P^{\DK}_1} \I \nonumber \\
			&& + \frac{2a^2r^2 \Delta \sin^2\theta}{P^{\DK}_1} \gamma_1 + \frac{\bar{\rho}^{2} (r^2+a^2) \Delta}{\rho^2 P^{\DK}_1} \Theta_3 - \frac{a}{M} \frac{\Delta\sin^2\theta}{\rho^2} \frac{P^{\DK}_3}{P^{\DK}_1} \gamma_3 + \frac{2\Delta}{\rho^2} \frac{P^{\DK}_4}{P^{\DK}_1} \gamma_4 \nonumber \\
			&& - \frac{a^2}{2M} \frac{\rho^2 \Delta^2 \sin^2\theta}{P^{\DK}_1} \frac{d\gamma_1}{dr} - \frac{a}{2M} \frac{\Delta^2 (\Sigma+2a^2Mr\sin^2\theta) \sin^2\theta}{P^{\DK}_1} \frac{d \gamma_3}{dr} - \frac{a^2}{2M} \frac{\Delta^2 (\rho^2-4Mr) \sin^2\theta}{P^{\DK}_1} \frac{d\gamma_4}{dr} \,, \nonumber \\
		h^{\DK}_{rr} &=& - \frac{1}{\Delta} \I -\frac{1}{\Delta} \Theta_3 \,,
\nonumber \\
	h^{\DK}_{\phi\phi} &=& -\frac{(r^2+a^2)^2}{a^2} h^{\DK}_{tt} + \frac{\Delta}{a^2} \I - \frac{2(r^2+a^2)}{a} h^{\DK}_{t\phi} + \frac{\Delta}{a^2} \Theta_3 - \frac{2\Delta^2 \sin^2\theta}{a} \gamma_3 + \frac{2\Delta^2}{a^2} \gamma_4 \,, \nonumber \\
	\frac{\partial h^{\DK}_{\theta\theta}}{\partial r} &=& \frac{2r}{\rho^2} h^{\DK}_{\theta\theta} + \frac{2a^2 \sin\theta \cos\theta}{\rho^2} h^{\DK}_{r\theta} + 2 \frac{\partial h^{\DK}_{r\theta}}{\partial \theta} + \frac{2r}{\rho^2} \I - 2r \, \gamma_1 + \frac{2r}{\rho^2} \Theta_3 \,, \nonumber \\
	\frac{\partial^2 h^{\DK}_{t\phi}}{\partial r^2} &=& \frac{8aM \sin\theta \cos\theta}{\rho^8} \frac{P^{\DK}_5}{P^{\DK}_1} h^{\DK}_{r\theta} - \frac{4aMr(r^2+a^2) \sin\theta \cos\theta}{\rho^6} \frac{\partial h^{\DK}_{r\theta}}{\partial r} + \frac{2a^2\sin^2\theta}{\rho^4} \frac{P^{\DK}_6}{P^{\DK}_1} h^{\DK}_{t\phi} - \frac{2r}{\rho^2}\frac{P^{\DK}_7}{P^{\DK}_1} h^{\DK}_{t\phi} \nonumber \\
		&& + \frac{4aMr\sin^2\theta}{\rho^4}\frac{P^{\DK}_{15}}{P^{\DK}_{16}} \I - \frac{4aMr\sin^2\theta}{\rho^4}\frac{P^{\DK}_8}{P^{\DK}_1} \gamma_1 + \frac{4aMr}{\rho^4}\frac{P^{\DK}_9}{P^{\DK}_1} \Theta_3 + \frac{2\sin^2\theta}{\rho^4}\frac{P^{\DK}_{10}}{P^{\DK}_1} \gamma_3 \nonumber \\
		&& - \frac{16aM \sin^2\theta}{\rho^4}\frac{P^{\DK}_{11}}{P^{\DK}_1} \gamma_4 - \frac{2a}{\rho^4}\frac{P^{\DK}_{12}}{P^{\DK}_1} \frac{d\gamma_1}{dr} - \frac{2\sin^2\theta}{\rho^4}\frac{P^{\DK}_{13}}{P^{\DK}_1} \frac{d\gamma_3}{dr} - \frac{2a\sin^2\theta}{\rho^4}\frac{P^{\DK}_{14}}{P^{\DK}_1} \frac{d\gamma_4}{dr} - \frac{a \Delta \sin^2\theta}{\rho^2} \frac{d^{2} \gamma_1}{dr^{2}} \nonumber \\
		&& - \frac{\Delta \sin^2\theta}{\rho^4}(\Sigma+2a^2Mr\sin^2\theta) \frac{d^{2} \gamma_3}{dr^{2}} - \frac{a \Delta (\rho^2-4Mr) \sin^2\theta}{\rho^4} \frac{d^{2} \gamma_4}{dr^{2}} \,,
		\label{Carter-conds-DK}
\ea
\end{widetext}
where recall that $\bar{\rho}^{2} \equiv r^{2} - a^{2} \cos^{2}{\theta}$, $\Theta_{r}=\Theta_{r}(\theta)$ is an arbitrary function of polar angle, while $P^{\DK}_i$ are polynomials in $r$ and $\cos\theta$, given explicitly in Appendix~\ref{app:pols}. Notice that many of these relations are integro-differential, as $\I$ is defined as
\ba
\I &=& \int dr \; \left[ \frac{2a^2 \sin\theta \cos\theta}{\rho^2} h^{\DK}_{r\theta} + 2r \, \gamma_1 + \rho^2 \, \frac{d \gamma_1}{dr} \right] \,.
\ea
Notice also that the component $h^{\DK}_{r\theta}$ is free and thus there are here truly only four independent metric components.

We see then that any metric perturbation $h_{\alpha \beta}^{\DK}$ that satisfies Eq.~\eqref{Carter-conds-DK} will possess a Carter constant. If given a specific non-Kerr metric, one can then use these equations to reconstruct the $\gamma_{i}$ and $\Theta_{2}$ functions to automatically obtain the perturbative second-order Killing tensor associated with this spacetime.

With this at hand, one can now construct the constants of the motion. The metric in Eqs.~\eqref{full-g2} and~\eqref{DK-hab} remains stationary and axisymmetric since all metric components $h_{\alpha \beta}^{\DK}$ depend only on $(r,\theta)$. Therefore, the constants of the motion can be expanded in exactly the same way as in Eqs.~\eqref{EqE}-\eqref{EqC}. With this at hand, the geodesic equations can be rewritten in first-order form in exactly the same way as in the BK case, except that here different metric components are non-vanishing. Of course, all perturbed quantities must here have a DK superscript.

\subsubsection{Non-Rotating Limit}
\label{DK-non-rot}

Let us now take the non-rotating limit, i.e.~$a \to 0$, of the DK Carter conditions. As before, we expand all arbitrary functions as in Eqs.~\eqref{low-a-exp-arb} and~\eqref{low-a-exp-g}.
Imposing the constraints
\ba
\Theta_{3,0} &=& 0\,,
\qquad
\gamma_{1,0} =  - h^{\DK}_{rr,0} f\,
\nonumber \\
\gamma_{3,0} &=& 0\,,
\qquad
\gamma_{3,1} = \frac{h_{rr,0}^{\DK}}{2 r^{2}} - \frac{r-4 M}{2 r^{3} f^{2}} h^{\DK}_{tt,0}\,
\nonumber \\
\gamma_{4,0} &=& \frac{h^{\DK}_{tt,0}}{2 f^{2}} + \frac{h^{\DK}_{rr,0}}{2}\,,
\ea 
forces the metric components to take the form
\ba
h^{\DK}_{tt} &=& h^{\DK}_{tt,0}\,,
\qquad 
h^{\DK}_{rr} = h^{\DK}_{rr,0}\,,
\qquad 
h^{\DK}_{\phi\phi} = 0\,.
\ea
All other pieces of ${\cal{O}}(a^{0})$ in the remaining metric components can be set to zero by setting $(h^{\DK}_{r\theta},h^{\DK}_{\theta\theta},h^{\DK}_{t\phi})$ to zero. With these choices, the differential conditions are automatically satisfied, where we have set all integration constants to zero.

The Killing tensor is then given by Eq.~\eqref{KT-parameterization}, with the perturbed null vector components
\ba
\delta k_{\DK}^{\alpha} &=& \left[- \frac{h^{\DK}_{rr,0}}{2} + \frac{h^{\DK}_{tt,0}}{2 f^{2}} , - h^{\DK}_{rr,0} f, 0, {\cal{O}}(a) \right]\,,
\\ \nonumber 
\delta l_{\DK}^{\alpha} &=& \left[ \frac{h^{\DK}_{rr,0}}{2} + \frac{h^{\DK}_{tt,0}}{2 f^{2}} , 0 ,0, {\cal{O}}(a) \right]\,,
\ea
where here we have set $\delta l_{\DK}^{r} = 0$. As before, the Killing tensor reduces exactly to that of the Schwarzschild spacetime, with the only non-zero components being $(\bar{\xi}_{\theta \theta}, \bar{\xi}_{\phi \phi}) = r^{4}(1, \sin^{2}{\theta})$.

\section{Relating Parameterizations}
\label{sec:Comp}

\subsection{To Each Other}

In the previous sections, we proposed two different parameterizations of deformed spacetimes suitable for modeling alternative theory predictions. These parameterizations are related via a gauge transformation. Under a general diffeomorphism, the metric transforms according to $h^{\DK}_{\mu\nu} \to h^{\DK'}_{\mu \nu} \equiv h^{\DK}_{\mu\nu} + \nabla_{(\mu}\xi_{\nu)}$, where we parameterize the generating vectors as
\ba
\xi = \left[\xi_0(r), \xi_1(r,\theta), \xi_2(r,\theta), \xi_3(r)\right] \,.
\label{gen-vec-form}
\ea
The question is whether a generating vector exists that could take a generic $h^{\DK}_{\mu \nu}$ metric perturbation to one of $h^{\BK}_{\mu \nu}$ form. The DK parameterization has an $(r,\theta)$ component that is absent in the BK one, while the BK one has $(t,r)$ and $(r,\phi)$ components that are absent in the DK parameterization. 

Let us assume that we have a certain metric in DK form. The first task is to remove the $(r,\theta)$ component, i.e.~to find a diffeomorphism such that $h^{\DK'}_{r \theta} = 0$.  This can be achieved by requiring that  
\be
h^{\DK}_{r\theta} + \frac{1}{2} \Sigma \left[ \frac{1}{\Delta} \frac{\partial\xi_1}{\partial\theta} +  \frac{\partial\xi_2}{\partial r} \right] =0\,,
\ee
whose solution is
\be
\xi_{1} = F_{1}(r) - \Delta \int \left(\frac{\partial \xi_{2}}{\partial r} + \frac{2 h^{\DK}_{r\theta}}{\rho^{2}} \right) d\theta\,,
\label{xi1-const}
\ee
where $F_{1}(r)$ is a free integration function and $\xi_{2}(r,\theta)$ is free. 

The generating vector of Eq.~\eqref{gen-vec-form} with the condition in Eq.~\eqref{xi1-const} not only guarantees that $h_{r \theta}^{\DK'} = 0$, but also ensures that the only non-vanishing components of the gauge-transformed metrics are the $(t,t)$, $(t,r)$, $(t,\phi)$, $(r,r)$, $(r,\phi)$, $(\theta,\theta)$ and $(\phi,\phi)$, exactly the same non-zero components as in the BK metric. The new components are 
\ba
h^{\DK'}_{tt} &=& h^{\DK}_{tt} - \frac{M \bar{\rho}^{2}}{\Sigma^2} \xi_1 + \frac{2a^2 M r \sin\theta \cos\theta}{\Sigma^2} \xi_2 \,,
\nonumber 
\\
h^{\DK'}_{tr} &=& - \frac{1}{2}\left(1-\frac{2Mr}{\Sigma}\right) \frac{d\xi_0}{dr} - \frac{aMr\sin^2\theta}{\Sigma} \frac{d\xi_3}{dr}\,,
\nonumber \\ 
h^{\DK'}_{t\phi} &=& h^{\DK}_{t\phi} + \frac{aM \bar{\rho}^{2} \sin^2\theta}{\Sigma^2} \xi_1 
\nonumber \\
&-& \frac{2aMr (r^2+a^2) \cos\theta \sin\theta}{\Sigma^2} \xi_2 \,,
\nonumber 
\\
h^{\DK'}_{rr} &=& h^{\DK}_{rr} + \frac{a^2r}{\Delta^2} \xi_1 - \frac{a^2 \sin\theta \cos\theta}{\Delta} \xi_2 + \frac{\Sigma}{\Delta} \frac{\partial\xi_1}{\partial r} \,,
\nonumber \\
h^{\DK'}_{r\phi} &=& - \frac{\sin^2\theta}{2 \Sigma} \left\{ 2aMr \frac{d\xi_0}{dr} 
+ \left[\left(r^2+a^2\right)^2
\right. \right. 
\nonumber \\
&+& \left. \left. 
a^2\Delta\sin^2\theta \right]\frac{d\xi_3}{dr} \right\} \,,
\nonumber 
\ea
\ba
h^{\DK'}_{\theta\theta} &=& h^{\DK}_{\theta\theta} + r \; \xi_1 - a^2 \sin\theta \cos\theta \; \xi_2 + \Sigma \; \frac{\partial \xi_2}{\partial \theta} \,,
 \nonumber \\  
h^{\DK'}_{\phi\phi} &=& h^{\DK}_{\phi\phi} + \frac{\sin^2\theta}{\Sigma^2} \left\{ r^5 - a^2Mr^2 
\right. 
\nonumber \\
&+& \left.
a^2\cos^2\theta \left[ 2r \left(r^2+M^2 \right) + \Delta M \right] 
\right.
 \nonumber \\
&+& \left. 
a^4\cos^4\theta  (r-M)  \right\} \xi_1 +\sin\theta \cos\theta  
\nonumber \\
&\times& 
\left[ \Delta + \frac{2Mr (r^2+a^2)^2}{\Sigma^2} \right] \xi_2\,,
\ea
where recall that $\bar{\rho}^{2} \equiv r^2-a^2\cos^2\theta$ and $\xi_{1}$ is given by Eq.~\eqref{xi1-const}.

The above result can be simplified somewhat by setting $\xi_{2} = 0$ and $F_{1}(r) = 0$ in Eq.~\eqref{xi1-const}. Notice that the $(t,r)$ and $(r,\phi)$ components are not modified by these vector components. The modified components then become
\ba
h^{\DK'}_{tt} &=& h^{\DK}_{tt} - \frac{M \bar{\rho}^{2}}{\Sigma^2} \xi_1\,,
\nonumber \\
h^{\DK'}_{t\phi} &=& h^{\DK}_{t\phi} + \frac{aM \bar{\rho}^{2} \sin^2\theta}{\Sigma^2} \xi_1 
\nonumber \\
h^{\DK'}_{rr} &=& h^{\DK}_{rr} + \frac{a^2r}{\Delta^2} \xi_1 + \frac{\Sigma}{\Delta} \frac{\partial\xi_1}{\partial r} \,,
\nonumber \\
h^{\DK'}_{\theta\theta} &=& h^{\DK}_{\theta\theta} + r \; \xi_1\,,
 \nonumber \\  
h^{\DK'}_{\phi\phi} &=& h^{\DK}_{\phi\phi} + \frac{\sin^2\theta}{\Sigma^2} \left\{ r^5 - a^2Mr^2 
\right. 
\nonumber \\
&+& \left.
a^2\cos^2\theta \left[ 2r \left(r^2+M^2 \right) + \Delta M \right] 
\right.
 \nonumber \\
&+& \left. 
a^4\cos^4\theta  (r-M)  \right\} \xi_1\,.
\ea

We have thus found a generic diffeomorphism that maps a DK metric to one that has the same non-zero components as a BK metric. Notice that $\xi_{0}$ and $\xi_{3}$ only enter to generate $(t,r)$ and $(r,\phi)$ components. If we know the form of the components that we are trying to map to, then we could solve for these two vector components. For example, let us assume that $h_{tr}^{\BK}$ and $h_{r \phi}^{\BK}$ are given and we wish to find $\xi_{0,3}$ such that $h_{tr}^{\DK'} = h_{tr}^{\BK}$ and $h_{r \phi}^{\DK'} = h_{r \phi}^{\BK}$. This implies 
\ba
0 &=& h^{\BK}_{tr} + \frac{1}{2}\left(1-\frac{2Mr}{\Sigma}\right) \frac{d\xi_0}{dr} - \frac{aMr\sin^2\theta}{\Sigma} \frac{d\xi_3}{dr}\,,
\\ \nonumber 
0 &=& h^{\BK}_{r\phi} + \frac{\sin^2\theta}{2 \Sigma} \left\{ 2aMr \frac{d\xi_0}{dr} 
\right. 
\nonumber \\
&+& \left. 
\left[\left(r^2+a^2\right)^2 +a^2\Delta\sin^2\theta \right]\frac{d\xi_3}{dr} \right\} \,.
\ea
Let us, for one moment, allow $\xi_{0,3}$ to be arbitrary functions of $(r,\theta)$. The
solution to the above system is then
\ba
\xi_{0} &=& F_{2}(\theta) + \frac{2}{\Delta} \int \frac{dr}{\rho^{2}} \left[
h_{tr}^{\BK} \cos^{2}{\theta} \left(2 a^{2} M r - a^{2} r^{2} - a^{4} \right) 
\right. 
\nonumber \\
&-& \left.
h_{tr}^{\BK} \left(2 a^{2} M r + a^{2} r^{2} + a^{4} \right) 
- 2 M r a h_{r\phi}^{\BK} \right]\,,
\\ \nonumber 
\xi_{3} &=& F_{3}(\theta) + \frac{2}{\Delta}  \int \frac{dr}{\rho^{2}}
\left[ h_{r\phi}^{\BK} \left( \rho^{2} - 2 M r \right) 
\right. 
\nonumber \\
&-& \left.
2 a M r \sin^{2}{\theta} h_{tr}^{\BK} \right]\,.
\ea
But of course, to avoid introducing other spurious components to the $\mathrm{DK}'$ metric, such as $h_{t\theta}^{\DK}$, we must require that $\xi_{0,3}$ be functions of $r$ only. This implies that the integrands must be themselves also functions of $r$ only. Setting the integrands equal to $F_{2}(r)$ and $F_{3}(r)$, we find the conditions
\ba
h_{tr}^{\BK} &=& \frac{F_{2}(r) \left(\rho^{2} - 2 M r \right) + 2 M r a \sin^{2}{\theta} F_{3}(r)}{\rho^{2}}\,,
\nonumber \\
h_{r \phi}^{\BK} &=& \frac{2 M r a  \sin^{2}{\theta} F_{2}(r)}{\rho^{2}} 
\nonumber \\
&-&
\frac{F_{3}(r)}{\rho^{2}}  \left[\rho^{4} + a^{4} \left(\rho^{2} + 2 M r \right)\right] \sin^{2}{\theta} \,.
\label{BK-conds}
\ea
Provided $h_{t r}^{\BK}$ and $h_{r \phi}^{\BK}$ components can be written in the above form, the generating function becomes
\be
\xi_{0} = - 2  \int F_{2}(r) dr\,, 
\qquad
\xi_{3} = - 2 \int F_{3}(r) dr\,.
\ee
%

\subsection{To Alternative Theories}
\subsubsection{Dynamical Chern-Simons Modified Gravity}

The BK and DK parameterizations can also be mapped to known alternative theory BH metrics. In dynamical CS gravity~\cite{Alexander:2009tp}, one can employ the slow-rotation approximation, where one assumes the BH's spin angular momentum is small, $|S|/M^{2} \ll 1$, and the small-coupling approximation, where one assumes the theory's corrections are small deformations away from GR to find an analytical BH solution. Yunes and Pretorius~\cite{Yunes:2009hc} found that this solution is simply $g_{\mu \nu} = \bar{g}_{\mu \nu} + h_{\mu \nu}^{\CS}$, where $\bar{g}_{\mu \nu}$ is the Kerr metric (expanded in the slow-rotation limit), while the only non-vanishing component of the deformation tensor in Boyer-Lindquist coordinates is 
\be
\label{CShtph}
h_{t \phi}^{\CS} = \frac{5}{8} \zeta_{\CS} \frac{a}{M} \frac{M^{5}}{r^{4}} \sin^{2}{\theta} \left(1 + \frac{12 M}{7 r} + \frac{27 M^{2}}{10 r^{2}} \right)\,,
\ee
where $a$ is the Kerr spin parameter, $M$ is the BH mass and $\zeta_{\CS}$ is a dimensionless constant that depends on the CS couplings and is assumed to be small.

Such a non-GR BH solution can be mapped to the generalized bumpy parameterization of Eqs.~\eqref{full-g} and \eqref{BK-metric-exp} via $\delta_2 =  \delta g_\phi^{\CS} = \delta_5$, and all other $\delta_{i}$ and $\Theta_{i}$ vanish. The quantity $\delta g_\phi^{\CS}$ is given by~\cite{2009PhRvD..80f4006S}
\ba
\delta g_\phi^{\CS} &=& - \zeta_{\CS} a M^{4} \frac{70r^2+120Mr+189M^2}{112r^8(1-2M/r)} \,.
\ea
With these choices, the full null vectors become
\ba
k^{\alpha} &=& \left[\frac{r^{2}+a^{2}}{\Delta},1,0,\frac{a}{\Delta} + \delta g_\phi^{\CS} \right]\,,
\nonumber \\
l^{\alpha} &=&  \left[\frac{r^{2}+a^{2}}{\Delta},-1,0,\frac{a}{\Delta} + \delta g_\phi^{\CS} \right]\,,
\label{pert-to-null-dir}
\ea
which agrees with Eq.~$(35)$ of~\cite{2009PhRvD..80f4006S}. Moreover, with these choices all the metric components vanish except for the $(t,\phi)$ one, which must be equal to Eq.~\eqref{CShtph} for the differential constraint in Eq.~\eqref{BK-metric-exp} to be satisfied.  

Similarly, we can also map it to the DK parameterization of Eqs.~\eqref{full-g2} and~\eqref{DK-hab} via $\gamma_{3} = \delta g_{\phi}^{\CS}$, and all other arbitrary functions vanish. With this, the perturbation to the principal null directions are equal to Eq.~\eqref{pert-to-null-dir}, which agrees with~\cite{2009PhRvD..80f4006S}. Moreover, we have checked that Eq.~\eqref{CShtph} satisfies the differential constraint found in Sec.~\ref{sec:General-Kerr}.

\subsubsection{Dynamical, Quadratic Gravity}

We can similarly map the BK and DK parameterizations to the BH solution found in dynamical quadratic gravity~\cite{Yunes:2011we}. Treating GR corrections as small deformations (i.e.~working in the small-coupling limit), Yunes and Stein~\cite{Yunes:2011we} found that the unique non-spinning BH solution is $g_{\mu \nu} = \bar{g}_{\mu \nu} + h_{\mu \nu}^{\QG}$, where $\bar{g}_{\mu \nu}$ is the Schwarzschild metric, while $h_{\mu \nu}^{\QG}$ in Schwarzschild coordinates is given by~\cite{Yunes:2011we}
\ba
h_{tt}^{\QG} &=& - \frac{\zeta_{\QG}}{3} \frac{M^{3}}{r^{3}} \left(1 +  \frac{26 M}{r} + \frac{66}{5} \frac{M^{2}}{r^{2}}
+ \frac{96}{5} \frac{M^{3}}{r^{3}} 
\right.
\nonumber \\
&-& \left. \frac{80 M^{4}}{r^{4}} \right)\,,
\nonumber \\
h_{rr}^{\QG} &=& -\frac{\zeta_{\QG}}{f^{2}} \frac{M^{2}}{r^{2}} \left(1 + \frac{M}{r}  + \frac{52}{3} \frac{M^{2}}{r^{2}} + \frac{2 M^{3}}{r^{3}} + \frac{16}{5} \frac{M^{4}}{r^{4}} 
\right. 
\nonumber \\
&-& \left. \frac{368}{3} \frac{M^{5}}{r^{5}} \right)\,,  
\ea
where $M$ is the mass of the Schwarzschild BH and $\zeta_{\QG}$ is a dimensionless constant that depends on the couplings of the theory and is assumed to be small

For a spherically symmetry background, one can easily show that the Schwarzschild Killing tensor (i.e.~one whose only non-vanishing components are $\xi_{\theta \theta} = r^{4}$ and $\xi_{\phi \phi} = r^{4} \sin^{2}{\theta}$) satisfies the Killing equation, regardless of the functional form of the $(t,t)$ and $(r,r)$ components of the metric. This means that the perturbed vectors $\delta k^{\alpha}$ and $\delta l^{\alpha}$ must adjust so that the above is true. We find that the following vector components do just that:
\ba
\delta k^{t} &=&  \frac{\zeta_{\QG} M^{2} r^{4}}{2 f r^{6}} \left(1 + \frac{8}{3} \frac{M}{r} 
+ \frac{14 M^{2}}{r^{2}} 
\right. 
\nonumber \\
&+& \left.
\frac{128}{5} \frac{M^{3}}{r^{3}} + \frac{48 M^{4}}{r^{4}} \right)\,, 
\nonumber \\
\delta k^{r} &=&  \frac{\zeta_{\QG} M^{2} r^{4}}{f r^{6}} \left(1 + \frac{M}{r}  + \frac{52}{3} \frac{M^{2}}{r^{2}}
 + \frac{2 M^{3}}{r^{3}} 
 \right.
 \nonumber \\
 &+& \left.
  \frac{16}{5} \frac{M^{4}}{r^{4}} 
 - \frac{368}{3} \frac{M^{5}}{r^{5}} \right)\,,
 \nonumber \\
\delta l^{t} &=& - \frac{\zeta_{\QG} M^{2} r^{4}}{2 f^{2} r^{6}} \left(1 + \frac{4}{3} \frac{M}{r} + \frac{26 M^{2}}{r^{2}}
+ \frac{32}{5} \frac{M^{3}}{r^{3}}  
\right. 
\nonumber \\
&+& \left.
 \frac{48}{5} \frac{M^{4}}{r^{4}}
-\frac{448}{3} \frac{M^{5}}{r^{5}}\right)\,, 
\ea
and all other components vanish.

That perturbed null vectors exist to reduce the Killing tensor to its Schwarzschild counterpart must surely be the case, as we have already shown in Sec.~\ref{BK-non-rot} and~\ref{DK-non-rot} that in the non-rotating limit both the BK and DK parameterizations allow for generic $(t,t)$ and $(r,r)$ deformations. 

\subsubsection{Arbitrary perturbations: perturbation to $\gamma_3$, perturbations to $\gamma_1$ and $\gamma_4$}
We can understand the functions that parametrize the metric perturbation by considering perturbations where we allow only a few of the parametrizing functions to be nonzero. Consider a perturbation in the deformed Kerr parametrization of Eqs.~\eqref{full-g2} and~\eqref{DK-hab} described by $\gamma_1 = \gamma_4 = \Theta_3 = 0$ and $h_{r\theta} = h_{\theta\theta} = 0$. In the small $a$ limit, the only nonzero component of the metric is the $(t,\phi)$ component:
\ba
h_{t\phi} = -f \left( r^2\sin^2\theta \, \gamma_3(r) + C \right) \,,
\ea
where $C$ is a constant. Now consider a perturbation in the DK parametrization whose only nonzero parameters are $\gamma_1$ and $\gamma_4$. In the small $a$ limit, this produces perturbations to the metric given by
\ba
h_{tt} &=& f \gamma_1(r) + 2 f^2 \gamma_4(r) \,, \nonumber \\
h_{rr} &=& -f^{-1} \gamma_1(r) \,, \nonumber \\
h_{\phi\phi} &=& f^2 \sin^2\theta \left[ \frac{r^4}{2M} \frac{d\gamma_1}{dr} + 2r^2 \gamma_4(r) \right. \nonumber \\
	&& \left. + \frac{r^3(r-4M)}{2M} \frac{d\gamma_4}{dr} \right] \,.
\ea
We then see that in the DK parameterization $\gamma_{3}$ controls modifications to frame-dragging (and thus the ergosphere), while $\gamma_{1}$ and $\gamma_4$ control modifications to the location of the event-horizon and the innermost-stable circular orbit. 

\section{Discussion}
\label{sec:Con}

The detection of GWs from EMRIs should allow for the detailed mapping of the spacetime metric on which the SCOs move. Such a mapping in turn allows for null tests of GR, as one would in principle be able to constrain deviations of the background spacetime from the Kerr metric. To carry out such tests, however, one requires a parameterization of the metric tensor that can handle model-independent, non-GR deviations. This was the purpose of this paper.  

We have constructed two such parameterizations: one as a generalization of the bumpy BH formalism (the BK scheme) and one as a generalized deformation of the Kerr metric (the DK scheme). The former promotes the non-vanishing components of the metric perturbation that contain bumps to arbitrary functions of radius and polar angle. The latter takes the most general stationary, axisymmetric line element without assuming Ricci-flatness and constructs a generic metric deformation from it. In both cases, the arbitrary functions introduced are constrained by requiring that there exists an approximate second-order Killing tensor.

These schemes differ from each other in the metric components that are assumed not to vanish. We have found a gauge transformation, however, that relates them. That is, we have shown that a generating vector exists such that the BK metric can be mapped to the DK one. We have also mapped both parameterizations to known analytical non-GR solutions, thus automatically finding their respective Killing tensors and Carter constants. 

The perturbative nature of our approach puts some limitation on the generality of the deformation. All throughout we restricted the metric deformation to be a small deviation away from the GR solution. But due to the structure of the perturbation, there can be regions in spacetime where this deformation dominates over the GR background. For example, the full metric tensor might lose Lorentzian signature if the metric perturbation dominates over the Kerr background and it possesses a certain definite sign, i.e., the metric will no longer have ${\rm{det}}(g) < 0$. Clearly, such regions are unphysical, and the coupling constants that control the magnitude of the perturbation should be adjusted appropriately to ensure that if they exist, they are hidden inside the horizon. 

We should emphasize that we were guided by two principles in the construction of the parameterized schemes proposed here. First, we wanted to ensure that the parameterizations would easily map to known analytic solutions. Second, we wished to retain a smooth Kerr limit, such that when the deformation parameters are taken to zero, the deformed metric goes smoothly to Kerr. This in turn guaranteed that certain properties of the Kerr background were retained, such as the existence of a horizon. We additionally required that the metric tensors allowed a second-order perturbative Killing tensor, so that there exists a perturbative Carter constant. This latter requirement is not strictly necessary to perform null tests, as one could build GWs from the evolution of the full second-order equations of motion without rewriting them in first-order form.   

An interesting avenue of future research would be to investigate the Petrov type~\cite{1969eisp.book.....P} of the BK and DK metrics. Brink has shown that GR spacetimes that admit a Carter constant must also be of Type D~\cite{Brink:2009mq}. In alternative theories of gravity, however, a formal mathematical proof of the previous statement is lacking. We have here found Ricci-curved metrics that admit a Carter constant. One could now attempt to prove that the perturbative Carter existence conditions found in this paper also automatically imply that the spacetime is approximately Petrov Type D. This is a purely mathematical result that is beyond the scope of this paper. 

An interesting physical question that could now be answered with either the DK or BK frameworks is whether LISA could place interesting constraints on non-GR Kerr deformations. To answer this question, one would have to evolve geodesics in either framework and construct the associated GWs. One could then carry out a Fisher analysis to determine the magnitude of the non-GR deviation from Kerr that could be measured or constrained by LISA. We will investigate this problem in a forthcoming publication.  

\acknowledgments 

We would like to thank Scott Hughes and Eric Poisson for useful discussions. NY acknowledges support from NASA through the Einstein Postdoctoral Fellowship Award Number PF0-110080 issued by the Chandra X-ray Observatory Center, which is operated by the Smithsonian Astrophysical Observatory for and on behalf of NASA under contract NAS8-03060. SV and LS acknowledge support from NSF Grant PHY-0449884 and NASA Grant NNX08AL42G. LS acknowledges support from MIT's Solomon Buchsbaum fund.

\bibliographystyle{apsrev}
\bibliography{phyjabb,master}

\appendix
\section{Polynomial Functions}
\label{app:pols}
We here provide explicit expressions for the polynomials appearing in the conditions that guarantee the existence of a Carter constant. In Eq.~\eqref{BK-metric-exp}, the polynomials $P_{i}^{\DK}$ are given by
\begin{widetext}
\ba
P^{\BK}_1 &=& r^6+5a^2r^4+2a^4r^2 - 2a^2\cos^2\theta(2r^4+a^2r^2+a^4) - a^4\cos^4\theta(r^2-a^2) \,,
\nonumber \\
P^{\BK}_2 &=& r^4M+a^2r^2(r+2M) - a^2\cos^2\theta(r^2(r+2M)-a^2(r-2M)) - a^4\cos^4\theta(r-M) \,,
\nonumber \\
P^{\BK}_3 &=& r(r^3+a^2(r+4M)) + a^2\cos^2\theta(r^2-4Mr+a^2) \,,
\nonumber \\
P^{\BK}_4 &=& r^3(r^2(r-2M)+a^2(3r-4M)) - 2a^2r^2\cos^2\theta(r^2-2Mr-a^2) - a^4\cos^4\theta(3r^2-2Mr+a^2) \,,
\nonumber \\
P^{\BK}_5 &=& r^3(r-2M)(r^2(r+2M)+a^2(r+4M)) - r^3\cos^2\theta(r^4-4M^2r^2-a^2(r^2-6Mr+16M^2)-2a^4) 
\nonumber \\
&& - a^2\cos^4\theta(2r^3(r^2-2Mr+4M^2)+a^2r(r^2+4Mr-4M^2)-a^4(r-2M)) - a^4\cos^6\theta(r-2M)\Delta \,,
\nonumber \\
P^{\BK}_{6} &=& -\frac{1}{2} {\frac {\Delta\, \left( {\rho}^{2}\Delta-2\,Mr{\rho}^{2}+4\,Mr{a
}^{2}+4\,M{r}^{3} \right)  \sin^{2} \theta}{{\rho}^{4}}}\,,
\qquad
P^{\BK}_{7} = \frac{1}{2} {\frac {a \, \Delta\, \left( \sin \left( \theta \right)  \right) ^{2}
 \left( -{\rho}^{2}+4\,Mr \right) }{{\rho}^{4}}}\,,
 \\
P^{\BK}_{8,n} &=& - \left( 8\,{r}^{11}-4\,{r}^{7}{\rho}^{4}+16\,{r}^{9}{M}^{2}+2\,{r}^{9
}{\rho}^{2}-44\,{a}^{6}M{\rho}^{2}\Delta - 8\,{a}^{10}M-68\,{a}^{4}M{r}^
{4}{\rho}^{2}+128\,{a}^{4}{M}^{2}{r}^{3}{\rho}^{2}-56\,{a}^{6}r{M}^{2}
{\rho}^{2}
\right. 
\nonumber \\
&-& \left.
12\,{a}^{4}{r}^{5}{\rho}^{2}-16\,{r}^{3}{a}^{6}{\rho}^{2}-6
\,{a}^{8}r{\rho}^{2}-36\,{a}^{2}M{r}^{6}{\rho}^{2}+40\,{a}^{2}M{r}^{4}
{\rho}^{4}+32\,{r}^{9}{a}^{2}-3\,{r}^{5}{\rho}^{6}-8\,{r}^{10}M-20\,{a
}^{2}r{M}^{2}{\rho}^{6}
\right. 
\nonumber \\
&-& \left.
16\,{r}^{8}M{a}^{2}-16\,{r}^{7}{a}^{2}{M}^{2}-
48\,{r}^{3}{a}^{6}{M}^{2}+16\,{r}^{4}M{a}^{6}-56\,{a}^{2}{M}^{2}{r}^{3
}{\rho}^{4}+64\,{a}^{2}{M}^{2}{r}^{5}{\rho}^{2}+72\,{a}^{4}r{M}^{2}{
\rho}^{4}+48\,{r}^{7}{a}^{4}
\right. 
\nonumber \\
&+& \left.
32\,{r}^{5}{a}^{6}+8\,{r}^{3}{a}^{8}+60\, {a}^{4}M{\rho}^{4}\Delta-18\,{a}^{2}M{\rho}^{6}\Delta-80\,{a}^{4}{M}^{
2}{r}^{5}+8\,{a}^{8}M\Delta+16\,{a}^{8}{M}^{2}r+36\,{a}^{8}M{\rho}^{2}
-32\,{r}^{7}{M}^{2}{\rho}^{2}
\right. 
\nonumber \\
&-& \left.
4\,{r}^{8}M{\rho}^{2} -12\,{r}^{3}{a}^{4} {\rho}^{4}-12\,{a}^{2}{r}^{5}{\rho}^{4}-4\,M{r}^{6}{\rho}^{4}-4\,{a}^{
6}r{\rho}^{4}-44\,{a}^{6}M{\rho}^{4}+12\,M{r}^{4}{\rho}^{6}-8\,{M}^{2}
{r}^{3}{\rho}^{6}+24\,{M}^{2}{r}^{5}{\rho}^{4}\right. 
\nonumber \\
&+& \left.
2\,{r}^{3}{a}^{2}{\rho}^{6}+5\,{a}^{4}r{\rho}^{6}+12\,{a}^{4}M{\rho}^{6} \right)  \sin^{2}\theta\,,
\nonumber \\
P^{\BK}_{8,d} &=& \left( 4\,{r}^{6}+2\,{\rho}^{4}{a}^{2}-4\,{a}^{2}{\rho}^{2}\Delta-8\,
{a}^{2}{\rho}^{2}Mr+2\,{a}^{4}{\rho}^{2}+8\,{r}^{4}{a}^{2}-{\rho}^{4}
\Delta-2\,{\rho}^{4}Mr-2\,{r}^{4}{\rho}^{2}+4\,{a}^{4}\Delta+8\,Mr{a}^
{4}-4\,{a}^{6} \right) {\rho}^{4}\,,
\nonumber \\
P^{\BK}_{9,n} &=& a \sin^{2} \theta  \left( -16\,{M}^{2}{
r}^{5}{\rho}^{2}+8\,{M}^{2}{r}^{3}{\rho}^{4}-4\,M{r}^{4}{\rho}^{4}-20
\,M{r}^{6}{\rho}^{2}-32\,{a}^{2}{M}^{2}{r}^{5}+24\,{r}^{8}M-24\,{a}^{4
}{r}^{5}-8\,{a}^{8}M-8\,{r}^{9}
\right.
\nonumber \\
&+& \left.
56\,{a}^{2}M{r}^{6}+40\,{a}^{4}M{r}^{4
}-24\,{r}^{7}{a}^{2}-8\,{r}^{3}{a}^{6}+8\,{a}^{6}M\Delta+16\,{a}^{6}{M
}^{2}r-12\,{a}^{4}{\rho}^{4}M+28\,{a}^{6}M{\rho}^{2}-36\,{a}^{4}M{\rho
}^{2}\Delta
\right.
\nonumber \\
&+& \left.
12\,{a}^{2}{\rho}^{4}M\Delta-48\,{a}^{4}{M}^{2}{r}^{3}+8\,
r{a}^{2}{M}^{2}{\rho}^{4}-3\,{a}^{2}r{\rho}^{6}+16\,{r}^{7}{M}^{2}+2\,
{a}^{2}M{\rho}^{6}+18\,{a}^{2}{r}^{5}{\rho}^{2}+6\,{r}^{7}{\rho}^{2}+
18\,{r}^{3}{a}^{4}{\rho}^{2}
\right.
\nonumber \\
&+& \left.
6\,{a}^{6}r{\rho}^{2}+{r}^{3}{\rho}^{6}+
48\,{a}^{2}{M}^{2}{r}^{3}{\rho}^{2}-40\,{a}^{4}{M}^{2}r{\rho}^{2}-48\,
{a}^{2}M{r}^{4}{\rho}^{2} \right)\,,
\nonumber \\
P^{\BK}_{9,d} &=& {\rho}^{4} \left( 4\,{r}^{6}+2\,{\rho}^{4}{a}^{2}-4\,{a}^{2}{\rho}^{2}
\Delta-8\,{a}^{2}{\rho}^{2}Mr+2\,{a}^{4}{\rho}^{2}+8\,{r}^{4}{a}^{2}-{
\rho}^{4}\Delta-2\,{\rho}^{4}Mr-2\,{r}^{4}{\rho}^{2}+4\,{a}^{4}\Delta+
8\,Mr{a}^{4}-4\,{a}^{6} \right)\,,
\nonumber \\
P^{\BK}_{10,n} &=& \sin^{2}\theta  \left( -12\,rM{a}^{4}
{\rho}^{4}-48\,{a}^{2}{r}^{8}+6\,{r}^{4}{a}^{2}{\rho}^{4}-48\,{a}^{4}{
r}^{6}+16\,{a}^{6}{M}^{2}\Delta+32\,{a}^{6}{M}^{3}r+8\,{a}^{6}{\rho}^{
2}\Delta-8\,{a}^{2}{M}^{2}{\rho}^{6}-16\,{r}^{10}
\right.
\nonumber \\
&-& \left.
16\,{M}^{2}{r}^{8}+6\,{a}^{2}{\rho}^{6}Mr+24\,{r}^{4}{a}^{4}{\rho}^{2}+32\,{r}^{9}M-32\,{a
}^{2}{\rho}^{4}M{r}^{3}-8\,{a}^{8}{\rho}^{2}+8\,{a}^{6}{M}^{2}{\rho}^{
2}+40\,{M}^{2}{r}^{6}{\rho}^{2}+16\,rM{a}^{6}{\rho}^{2}
\right.
\nonumber \\
&-& \left.
32\,{r}^{5}M{a}^{2}{\rho}^{2}+8\,{r}^{8}{\rho}^{2}-16\,{a}^{8}{M}^{2}+96\,{a}^{2}{M}
^{2}{r}^{4}{\rho}^{2}-16\,{r}^{4}{a}^{6}+32\,{r}^{5}M{a}^{4}-80\,{r}^{
6}{a}^{2}{M}^{2}-48\,{r}^{4}{a}^{4}{M}^{2}+16\,{r}^{3}M{a}^{4}{\rho}^{
2}
\right.
\nonumber \\
&+& \left.
24\,{a}^{2}{r}^{6}{\rho}^{2}-6\,M{r}^{3}{\rho}^{6}+3\,{r}^{4}{\rho}
^{6}+2\,{r}^{6}{\rho}^{4}+5\,{a}^{4}{\rho}^{6}+64\,{r}^{7}M{a}^{2}+24
\,M{r}^{5}{\rho}^{4}-24\,{M}^{2}{r}^{4}{\rho}^{4}-48\,M{r}^{7}{\rho}^{
2}
\right.
\nonumber \\
&-& \left.
24\,{a}^{6}{M}^{2}{\rho}^{2} \sin^{2}\theta -6\,{a}^{2}{\rho}^{6}\Delta+6\,{a}^{4}{\rho}^{4}\Delta-4
\,{a}^{6}{\rho}^{4} \right)\,,
\nonumber \\
P^{\BK}_{10,d} &=& {\rho}^{4} \left( 4\,{r}^{6}+2\,{\rho}^{4}{a}^{2}-4\,{a}^{2}{\rho}^{2}
\Delta-8\,{a}^{2}{\rho}^{2}Mr+2\,{a}^{4}{\rho}^{2}+8\,{r}^{4}{a}^{2}-{
\rho}^{4}\Delta-2\,{\rho}^{4}Mr-2\,{r}^{4}{\rho}^{2}+4\,{a}^{4}\Delta+
8\,Mr{a}^{4}-4\,{a}^{6} \right)\,,
\nonumber \\
P^{\BK}_{11,n} &=& -8\,aM \sin^{2}\theta \left( {r}^{3}{
\rho}^{4}-5\,{\rho}^{2}{r}^{5}+4\,{r}^{7}-3\,r{a}^{2}{\rho}^{4}+6\,{a}
^{2}{r}^{3}{\rho}^{2}+{a}^{2}{\rho}^{4}M+{a}^{2}{\rho}^{2}M\Delta+2\,{
a}^{2}{\rho}^{2}{M}^{2}r-3\,{a}^{4}M{\rho}^{2}-8\,{a}^{2}M{r}^{4}
\right.
\nonumber \\
&+& \left.
3\,{a}^{4}r{\rho}^{2}-2\,M{r}^{6}-4\,{a}^{4}{r}^{3}+3\,M{r}^{4}{\rho}^{2}+
2\,{a}^{4}M\Delta+4\,{a}^{4}{M}^{2}r-2\,{a}^{6}M \right)\,,
\nonumber \\
P^{\BK}_{11,d} &=& {\rho}^{4} \left( 4\,{r}^{6}+2\,{\rho}^{4}{a}^{2}-4\,{a}^{2}{\rho}^{2}
\Delta-8\,{a}^{2}{\rho}^{2}Mr+2\,{a}^{4}{\rho}^{2}+8\,{r}^{4}{a}^{2}-{
\rho}^{4}\Delta-2\,{\rho}^{4}Mr-2\,{r}^{4}{\rho}^{2}+4\,{a}^{4}\Delta+
8\,Mr{a}^{4}-4\,{a}^{6} \right)\,,
\nonumber \\
P^{\BK}_{12,n} &=&-2\,r \left( -28\,{a}^{2}{\rho}^{2}\Delta-56\,{a}^{2}{\rho}^{2}Mr+18\,
{a}^{4}{\rho}^{2}+{\rho}^{6}+2\,{\rho}^{4}\Delta+4\,{\rho}^{4}Mr+4\,{
\rho}^{4}{a}^{2}-18\,{r}^{4}{\rho}^{2}+16\,{r}^{6}
\right.
\nonumber \\
&+& \left.
32\,{r}^{4}{a}^{2}+16\,{a}^{4}\Delta+32\,Mr{a}^{4}-16\,{a}^{6} \right)\,,
\nonumber \\
P^{\BK}_{12,d} &=& {\rho}^{2} \left( 4\,{r}^{6}+2\,{\rho}^{4}{a}^{2}-4\,{a}^{2}{\rho}^{2}
\Delta-8\,{a}^{2}{\rho}^{2}Mr+2\,{a}^{4}{\rho}^{2}+8\,{r}^{4}{a}^{2}-{
\rho}^{4}\Delta-2\,{\rho}^{4}Mr-2\,{r}^{4}{\rho}^{2}+4\,{a}^{4}\Delta+
8\,Mr{a}^{4}-4\,{a}^{6} \right)\,,
\nonumber \\
P^{\BK}_{13,n} &=& 2\,{a}^{2} \left( \sin \left( \theta \right)  \right) ^{2} \left( 8\,{
a}^{2}{\rho}^{2}\Delta+16\,{a}^{2}{\rho}^{2}Mr-8\,{a}^{4}{\rho}^{2}-{
\rho}^{6}-6\,{\rho}^{4}\Delta-12\,{\rho}^{4}Mr+8\,{\rho}^{4}{a}^{2}+24
\,{r}^{4}{\rho}^{2}-16\,{r}^{6}-16\,{r}^{4}{a}^{2} \right) \,,
\nonumber \\
P^{\BK}_{13,d} &=& {\rho}^{4} \left( 4\,{r}^{6}+2\,{\rho}^{4}{a}^{2}-4\,{a}^{2}{\rho}^{2}
\Delta-8\,{a}^{2}{\rho}^{2}Mr+2\,{a}^{4}{\rho}^{2}+8\,{r}^{4}{a}^{2}-{
\rho}^{4}\Delta-2\,{\rho}^{4}Mr-2\,{r}^{4}{\rho}^{2}+4\,{a}^{4}\Delta+
8\,Mr{a}^{4}-4\,{a}^{6} \right)\,,
\nonumber  \\ \nonumber 
P^{\BK}_{14} &=& {\frac {-4\, aMr \left( \sin \left( \theta \right)  \right) ^{2}
 \left( -4\,{r}^{4}+{\rho}^{2}\Delta+2\,{\rho}^{2}Mr-4\,{a}^{2}{\rho}^
{2}+4\,{r}^{2}{a}^{2} \right) }{{\rho}^{4} \left( 4\,{r}^{6}+2\,{\rho}
^{4}{a}^{2}-4\,{a}^{2}{\rho}^{2}\Delta-8\,{a}^{2}{\rho}^{2}Mr+2\,{a}^{
4}{\rho}^{2}+8\,{r}^{4}{a}^{2}-{\rho}^{4}\Delta-2\,{\rho}^{4}Mr-2\,{r}
^{4}{\rho}^{2}+4\,{a}^{4}\Delta+8\,Mr{a}^{4}-4\,{a}^{6} \right) }}\,.
\ea
\end{widetext}

Similarly, the polynomial functions that appear in the Carter conditions for the DK metric of Eq.~\eqref{Carter-conds-DK} are
\begin{widetext}
\begin{eqnarray}
	P^{\DK}_1 &=& r^2(r^4+5a^2+r^2+2a^4) - 2a^2\cos^2\theta (2r^4+a^2r^2+a^4) - a^4\cos^4\theta (r^2-a^2) \nonumber \,, \\
	P^{\DK}_2 &=& r^2(r^2M+a^2r+2a^2M) - a^2\cos^2\theta (r^2(r+2M)-a^2(r-2M)) - a^4\cos^4\theta (r-M) \,, \nonumber \\
	P^{\DK}_3 &=& r^3(r-2M)(r^2(r+2M)+a^2(r+4M)) +2r^3a^2\cos^2\theta (r^2-2Mr+4M^2+a^2) \nonumber \\
		&& + a^4\cos^4\theta (r-2M) \Delta \,, \nonumber \\
	P^{\DK}_4 &=& r^3(r^2(r-2M)+a^2(3r-4M)) - 2r^2a^2\cos^2\theta (r^2-2Mr-a^2) -a^4\cos^4\theta (3r^2-2Mr+a^2) \,, \nonumber \\
	P^{\DK}_5 &=& r^4(r^6+3a^2r^4+8a^4r^2+2a^6) - 3a^4r^4\cos^2\theta (3r^2-a^2) + a^4\cos^4\theta (5r^6-3a^2r^4+6a^4r^2+2a^6) \nonumber \\
		&& + a^6\cos^6\theta (2r^4-3a^2r^2-a^4) \,, \nonumber \\
	P^{\DK}_6 &=& r^4(r^2-6a^2) + 3a^2r^2 \cos^2\theta (3r^2+4a^2) - a^4\cos^2\theta (9r^2-2a^2) - a^6\cos^6\theta \,, \nonumber \\
	P^{\DK}_7 &=& r^6+10a^2r^4+6a^4r^2 - a^2\cos^2\theta (11r^4+16a^2r^2+10a^4) + a^4\cos^4\theta (5r^2+6a^2) + a^6\cos^6\theta \,, \nonumber \\
	P^{\DK}_8 &=& r^2(3r^2-a^2) - a^2\cos^2\theta (r^2-3a^2) \,, \nonumber \\
	P^{\DK}_9 &=& r^2(3r^4+5a^2r^2-2a^4) - 2a^2\cos^2\theta (r^2+a^2)(2r^2-3a^2) + a^4\cos^4\theta (r^2-3a^2) \,, \nonumber \\
	P^{\DK}_{10} &=& r^2(3r^2-a^2) - 3\cos^2\theta (r^4-a^4) + a^2\cos^4\theta (r^2-3a^2) \,, \nonumber \\
	P^{\DK}_{11} &=& -r^4(3r^6-2Mr^5+24a^2r^4-18a^2Mr^3-8a^2M^2r^2+19a^4r^2-24a^4Mr+48a^4M^2+6a^6) \nonumber \\
		&& + a^2\cos^2\theta (21r^8-18Mr^7-8M^2r^6+18a^2r^6-42a^2Mr^5+72a^2M^2r^4+33a^4r^4-32a^4Mr^3+24a^4M^2r^2 \nonumber \\
		&& +12a^6r^2-16a^6M^2) + a^4\cos^4\theta (11r^6+6Mr^5-24M^2r^4-12a^2r^4+22a^2Mr^3-24a^2M^2r^2+3a^4r^2 \nonumber \\
		&& -24a^4Mr+24a^4M^2+2a^6) + a^6\cos^6\theta (3r^4-6Mr^3-6a^2r^2+18a^2Mr-8a^2M^2-a^4) \,, \nonumber \\
	P^{\DK}_{12} &=& r^3(r^3M+3a^2r^2-6a^2Mr-a^4) - a^2\cos^2\theta (3r^5-3Mr^4-3a^2Mr^2-3a^4r+2a^4M) \nonumber \\
		&& + a^4\cos^4\theta (r^3-3a^2r+a^2M) \,, \nonumber \\
	P^{\DK}_{13} &=& r^3(r^6+9a^2r^4-10a^2Mr^3+6a^4r^2-8a^4Mr+2a^6) \nonumber \\
		&& - \cos^2\theta r(r^8+18a^2r^6-22a^2Mr^5+15a^4r^4-26a^4Mr^3+20a^6r^2-12a^6Mr+6a^8) \nonumber \\
		&& + a^2\cos^4\theta (9r^7-12Mr^6+6a^2r^5-22a^2Mr^4+27a^4r^3-18a^4Mr^2+6a^6r+4a^6M) \nonumber \\
		&& + a^4\cos^6\theta (3r^5+4Mr^4-10a^2r^3+6a^2Mr^2+3a^4r-6a^4M) + a^6 \cos^8\theta (r^3-3a^2r+2a^2M) \,, \nonumber \\
	P^{\DK}_{14} &=& r^3(3r^8-4Mr^7+22a^2r^6-30a^2Mr^5+8a^2M^2r^4+29a^4r^4-14a^4Mr^3+12a^6r^2-12a^6Mr-16a^6M^2+2a^8) \nonumber \\
		&& - a^2\cos^2\theta (15r^9-24Mr^8+8M^2r^7+10a^2Mr^6+21a^4r^5-34a^4Mr^4-32a^4M^2r^3+24a^6r^3+12a^6Mr^2 \nonumber \\
		&& -32a^6M^2r+6a^8r+8a^8M) - a^4\cos^4\theta (13r^7-32Mr^6+6a^2r^5+14a^2Mr^4+8a^2M^2r^3-3a^4r^3-42a^4Mr^2 \nonumber \\
		&& +48a^4M^2r+4a^6r-16a^6M) - a^6\cos^6\theta (3r^5-12Mr^4+8M^2r^3-2a^2r^3+18a^2Mr^2-16a^2M^2r-5a^4r+6a^4M) \,, \nonumber \\
	P^{\DK}_{15} &=& r^3(r^6-8M^2r^4+9a^2r^4-22a^2Mr^3+6a^4r^2-4a^4Mr+16a^4M^2+2a^6) \nonumber \\
		&& - a^2\cos^2\theta (9r^7-28Mr^6+9a^2r^5-18a^2Mr^4+16a^2M^2r^3+16a^2M^2r^3+18a^4r^3-36a^4Mr^2+6a^6r-8a^6M) \nonumber \\
		&& - a^4\cos^4\theta r(3r^4-4Mr^3+r^2(8M^2-9a^2)+18a^2Mr-16a^2M^2) - a^6\cos^6\theta (r^3-3a^2r+2a^2M) \,.
\end{eqnarray}
\end{widetext}
%

\section{Equations of Motion in Alternative Theories}
\label{alt-theories-EOM}
 
The derivation of the equations of motion in Sec.~\ref{subsec:bggeom-eom} is fairly general. In particular, this derivation is independent of the metric used; in no place did we use that the spacetime was Kerr. We required the divergence of the test particle stress-energy tensor vanishes; this condition is always true in GR, because of a combination of local stress-energy conservation and the equivalence principle. 

One might wonder whether the divergence-free condition of the stress-energy tensor holds in more general theories. In any metric GR deformation, the field equations will take the form
\be
G_{\alpha \beta} + {\cal{H}}_{\alpha \beta} = T_{\alpha \beta}^{\MAT} + T_{\alpha \beta}^{{\cal{H}}}\,,
\ee
where $G_{\alpha \beta}$ is the Einstein tensor, ${\cal{H}}_{\alpha \beta}$ is a tensorial deformation of the Einstein equations, and $T_{\alpha \beta}^{{\cal{H}}}$ is a possible stress-energy modification, associated with additional fields. The divergence of this equation then leads to
\be
\nabla^{\alpha} {\cal{H}}_{\alpha \beta} = \nabla^{\alpha} T_{\alpha \beta}^{\MAT} + \nabla^{\alpha} T_{\alpha \beta}^{{\cal{H}}}\,,
\ee
since the Bianchi identities force the divergence of the Einstein tensor to vanish. The Bianchi identities hold in alternative theories, as this is a geometric constraint and not one that derives from the action. We see then that the divergence of the matter stress-energy tensor vanishes independently provided  
\be
\nabla^{\alpha} {\cal{H}}_{\alpha \beta} = \nabla^{\alpha} T_{\alpha \beta}^{{\cal{H}}}\,.
\label{condition}
\ee

Whether this condition [Eq.~\eqref{condition}] is satisfied depends somewhat on the theory of interest. Theories that include additional degrees of freedom that couple both to the geometry and have their own dynamics usually satisfy Eq.~\eqref{condition}. This is because additional equations of motion arise upon variation of the action with respect to these additional degrees of freedom, and these additional equations reduce to Eq.~\eqref{condition}. Such is the case, for example, in dynamical CS modified gravity~\cite{2009PhRvD..80f4006S}. If no additional degrees of freedom are introduced, the satisfaction of Eq.~\eqref{condition} depends on whether the divergence of the new tensor ${\cal{H}}_{\alpha \beta}$ vanishes, which need not in general be the case. 

Recently, it was shown that the equations of motion are geodesic to leading order in the mass-ratio for any classical field theory that satisfies the following constraints~\cite{2010PhRvD..81h4060G}:
\begin{itemize}
\item It derives from a diffeomorphism-covariant Lagrangian, ensuring a Bianchi identity;
\item It leads to second-order field equations.
\end{itemize}
The second condition seems somewhat too stringent, as we know of examples where third-order field equations still lead to geodesic motion, i.e.~dynamical CS gravity~\cite{2009PhRvD..80f4006S}. Therefore, it seems reasonable to assume that this condition could be relaxed in the future. Based on this, we take the viewpoint that the equations of motion are geodesic even in the class of alternative theories we consider here. 

\end{document}